\RequirePackage{fix-cm}
\documentclass[onecolumn]{svjour3}
\smartqed  

\AtBeginDocument{%
  }

\journalname{Empirical Software Engineering}

\usepackage{amsmath}
\usepackage{booktabs} 
\usepackage[utf8]{inputenc}
\usepackage{float}
\usepackage{ifthen}
\usepackage{cite}
\usepackage{multirow}
\usepackage{amssymb,amsfonts}
\usepackage{caption} 
\usepackage[font=footnotesize]{caption, subcaption}
\usepackage{listings}
\usepackage{hyperref}
\hypersetup{colorlinks=true}
\usepackage{url,moreverb,xspace}
\usepackage{tcolorbox}
\tcbuselibrary{listings,breakable}
\usepackage{enumitem}
\usepackage{array,graphicx}
\usepackage{soul}
\usepackage{balance}
\usepackage{pifont}
\usepackage{nicefrac}
\usepackage{mathtools}
\usepackage{diagbox}
\usepackage{orcidlink} 
\usepackage{subcaption}
\usepackage{svg}

\usepackage[lined,boxruled,norelsize,linesnumbered]{algorithm2e}
\def\HiLi{\leavevmode\rlap{\hbox to \hsize{\color{gray!35}\leaders\hrule height .8\baselineskip depth .5ex\hfill}}}

\usepackage{graphicx}
\usepackage{seqsplit}

\newcommand{\xaiSG}{SmoothGrad\xspace} %
\newcommand{\xaiGC}{Grad-CAM++\xspace} %
\newcommand{\xaiIG}{Integrated Gradients\xspace} %
\newcommand{\xaiLime}{LIME\xspace} %
\newcommand{\tool}{\textsc{XMutant}\xspace} %
\newcommand{\deepjanus}{\text{DeepJanus}\xspace} %
\newcommand{\davetwo}{\mbox{DAVE-2}\xspace} %
\newcommand{\rom}[1]{\uppercase\expandafter{\romannumeral #1\relax}}

\SetKwComment{Comment}{$\triangleright$\ }{}
\SetCommentSty{itshape}
\newboolean{showcomments}
\setboolean{showcomments}{true}
  
\ifthenelse{\boolean{showcomments}}
    {\newcommand{\nb}[2] {
      \fcolorbox{black}{gray!20}{\bfseries\sffamily\scriptsize#1:}
      {\sf\small$\blacktriangleright$\textit{#2}$\blacktriangleleft$}
    }
    }
    {\newcommand{\nb}[2]{}
    }

\newcommand{\head}[1]{\noindent\textbf{#1.}}

\newcommand{\changed}[1]{\textcolor{black}{#1}}

\newcommand\mnist{MNIST\xspace}

\newcommand\imdb{IMDB\xspace}

\begin{document}



\title{\tool: XAI-based Fuzzing for Deep Learning Systems}

\author{
    Xingcheng Chen\textsuperscript{1,2} \orcidlink{0009-0002-0861-4093} \and
    Matteo Biagiola\textsuperscript{3} \orcidlink{0000-0002-7825-3409} \and
    Vincenzo Riccio\textsuperscript{4} \orcidlink{0000-0002-6229-8231} \and
    Marcelo d'Amorim\textsuperscript{5} \orcidlink{0000-0002-1323-8769} \and
    Andrea Stocco\textsuperscript{1,2} \orcidlink{0000-0001-8956-3894} 
}

\institute{
    \textsuperscript{1} Technical University of Munich, Munich, Germany. Email: xingcheng.chen@tum.de (corresponding author), andrea.stocco@tum.de \\
    \textsuperscript{2} fortiss GmbH, Munich, Germany. Email: xchen@fortiss.org, stocco@fortiss.org \\
    \textsuperscript{3} University of St. Gallen, Universit\`a della Svizzera italiana, St. Gallen/Lugano, Switzerland. Email: matteo.biagiola@{unisg,usi}.ch \\
    \textsuperscript{4} Universit\`a degli Studi di Udine, Udine, Italy. Email: vincenzo.riccio@uniud.it \\
    \textsuperscript{5} North Carolina State University, Raleigh, USA. Email: mdamori@ncsu.edu
}

\authorrunning{Chen et al.}
\titlerunning{\tool: XAI-based Fuzzing for Deep Learning Systems}




\date{Received: date / Accepted: date
}

\maketitle

\begin{abstract}
Semantic-based test generators are widely used to produce failure-inducing inputs for Deep Learning (DL) systems. They typically generate challenging test inputs by applying random perturbations to input semantic concepts until a failure is found or a timeout is reached. However, such randomness may hinder them from efficiently achieving their goal. This paper proposes \tool, a technique that leverages explainable artificial intelligence (XAI) techniques to generate challenging test inputs. \tool uses the local explanation of the input to inform the fuzz testing process and effectively guide it toward failures of the DL system under test. 
We evaluated different configurations of \tool in triggering failures for different DL systems both for model-level (sentiment analysis, digit recognition) and system-level testing (advanced driving assistance).
Our studies showed that \tool enables more effective and efficient test generation by focusing on the most impactful parts of the input. \tool generates up to $125\%$ more failure-inducing inputs compared to an existing baseline, up to 7$\times$ faster. We also assessed the validity of these inputs, maintaining a validation rate above {$89\%$}, according to automated and human validators.
\end{abstract}

\noindent \textbf{Keywords.}
Software testing, testing deep learning systems, explainable AI

\section{Introduction}\label{sec:introduction}

Deep Learning (DL) systems play a crucial role in software engineering~\cite{schutze2008introduction} due to their ability to solve complex tasks by learning from a corpus of data. However, DL systems pose additional challenges to testing. Unlike traditional software, where behavior can be anticipated through source code analysis, the actions of DL systems are less predictable~\cite{2020-Riccio-EMSE}. Therefore, test generation for DL systems is critical for ensuring their reliability and correctness. This process involves creating a diverse set of input data to uncover potential weaknesses or biases in the model, such as incorrect predictions or failures to generalize from data samples that are different from those available during training. Testing is particularly important in applications where DL is used for decision-making in critical scenarios, such as autonomous driving~\cite{2020-Stocco-ICSE,Abdessalem-ICSE18,Gambi:2019:ATS:3293882.3330566,survey-lei-ma}. 
Exhaustive testing of DL systems is impractical because of the size of the input space. Thus, generating effective test cases for these systems is an important problem, which requires a comprehensive understanding of the DL model's architecture, its training data, and the application domain.

Prior work proposed solutions to generate test cases automatically for DL systems~\cite{2020-Riccio-FSE,Abdessalem-ASE18-1,Abdessalem-ASE18-2,Abdessalem-ICSE18,9712397,Guannam_ETAL_FSE22,MULLINS2018197,Gambi:2019:ATS:3293882.3330566,drivefuzz,zhongETAL2021,liETAL2020,Jha2019MLBasedFI,10.1145/3597926.3598072}.
Some approaches such as DeepXplore~\cite{pei2017deepxplore}, DLFuzz~\cite{guo2018dlfuzz} and DeepTest~\cite{deeptest} target DL image classification systems and they involve raw input manipulation techniques that modify/corrupt pixel values of an input. These techniques do not generate new functional inputs as they produce minimal, often imperceptible changes to the original inputs, and are therefore suitable to test robustness and security deficiencies of the DL system~\cite{2023-Riccio-ICSE,2025-Maryam-ICST}. In contrast, functional test generation focuses on creating new inputs that deviate significantly from the original training distribution. These inputs target the long-tail problem of DL testing~\cite{zhang2024systematicreviewlongtailedlearning}, testing the DNN's ability to generalize to novel, unseen scenarios. Instances of functional test generators are the model-based approaches (also called semantic-based approaches) like \deepjanus~\cite{2020-Riccio-FSE}, DeepHyperion~\cite{zohdinasab2021deephyperion} and DeepMetis~\cite{2021-Riccio-ASE} or latent space manipulation techniques like SINVAD~\cite{kang2020sinvad,sinvad-tosem} and CIT4DNN~\cite{dola2024cit4dnn}.
They generate new inputs with mutations that are randomly applied to a semantic representation of the inputs~\cite{2020-Riccio-FSE}, or to a latent vector~\cite{kang2020sinvad}. While random mutations can eventually produce inputs that expose failures, these solutions are extraneous to the internal state of the DL system. Indeed, finding failures with black-box approaches is time-consuming, especially in computationally demanding contexts, such as simulation-based testing of self-driving cars.

This paper investigates the development of more effective and efficient semantic-based test generators for DL systems, leveraging eXplainable Artificial Intelligence (XAI) techniques. While methods similar to XAI, leveraging neuron activation values and gradient information, have been explored for raw input manipulation~\cite{neural_coverage_new,Ma:2018:DMT:3238147.3238202,Xie-ISSTA-2019} to derive adversarial attacks for robustness testing, their application to semantic-based functional test generation remains unexplored.
Therefore, we focus on semantic-based approaches that utilize an input model, as they have been successfully applied to various DL systems (e.g., classification, regression) and input types (e.g., text, images, logical driving scenarios).

In this work, we leverage the post-hoc local explanations of DL systems for individual predictions. Local explanations reveal the contributions of features in the input with respect to the model prediction~\cite{du2019techniques}. 
Depending on the model's input, the explanations can take the form of feature contributions (e.g., for textual inputs), or visual heatmaps (e.g., for image inputs)~\cite{jetley2018learn,samek2019explainable,samek2017explainable}.
Previous work highlights the valuable insights provided by XAI, particularly for understanding and debugging DL systems~\cite{abs-2201-00009,2022-Stocco-ASE,arxiv.2204.00480,DBLP:journals/corr/abs-2002-00863}. In our work, we leverage their information for guiding test generation by introducing \tool, a DL testing technique that leverages XAI's explanations to derive challenging inputs.
More specifically, \tool leverages a novel mutation operator that uses the local explanation on a given input to identify the area of its semantic representation that has higher contributions to the decision-making process of DL systems.
Our work shows that fuzzing such attention areas with targeted minor modifications accelerates fault exposure and preserves validity (i.e., the realism of the inputs) and label (i.e., the oracle).
\tool uses the local explanations in two ways. First, it uses them to rank and select the candidate semantic concepts from the semantic representation of input for mutation. Second, it uses these concepts to direct the mutation in the area of most attention of the DL system, enabling the approach to perform targeted input modifications.

We have evaluated \tool on three DL systems, representative of diverse DL tasks, namely sentiment analysis, digit recognition, and advanced driving assistance. These case studies differ in the forms of inputs (text, images, logical scenarios) and testing levels (model-level and system-level), resulting in different semantic-based input representations and XAI explanations.
In our experiments, accounting for more than 2$k$ test cases, \tool shows superior effectiveness and efficiency compared to \deepjanus, a state-of-the-art semantic-based test generator that has been successfully applied to multiple domains, yet its mutation operators are not aided by any guidance.
The guidance of the local explanations allows \tool to expose a higher number of failure-inducing inputs (up to $+208\%$ for sentiment analysis, $+125\%$ for digit recognition and $+27\%$ for advanced driving assistance) within half the iteration budget. 
Moreover, on average, \tool is also faster at exposing failure-inducing inputs for all DL systems (up to 3$\times$ times faster for sentiment analysis, 7$\times$ times faster for digit recognition and 2$\times$ times faster for advanced driving assistance). 
Our study also reveals that the failure-inducing inputs by \tool are valid in-distribution inputs, according to state-of-the-art automated input validators, and that they exhibit a high validity rate~($\approx$90\%) and label preservation rate~($\approx$70\%), for human assessors. 
Additionally, our evaluation revealed that gradient-guided raw input manipulation methods, despite utilizing neuron activation values and gradient information, only produce corrupted inputs. In contrast, \tool produces more natural failure-inducing inputs that remain within the original data manifold in semantic space.

Our paper makes the following contributions:

\begin{description}[noitemsep]
\item [\bf Approach.] A mutation operator for semantic-based fuzzing of DL systems based on local explanations that XAI techniques produce.
\item [\bf Technique.] An implementation of our approach for focused fuzzing of DL systems, implemented in the publicly available tool \tool~\cite{replication-package}. To the best of our knowledge, this is the first solution that uses XAI techniques for semantic-based focused test generation, both at the model and system level. 
\item [\bf Evaluation.] An empirical study on three different DL systems showing that \tool is better than a state-of-the-art approach, in terms of higher effectiveness, efficiency, validity, and label-preservation rates.
\end{description}
\section{Background}\label{sec:background}

Deep Neural Networks (DNNs) are increasingly used in complex safety-critical tasks, such as autonomous driving~\cite{nvidia-dave2}, autonomous aviation~\cite{Julian}, medical diagnosis~\cite{zhang2018medical} or disease prediction~\cite{Zhao-nature}. This paper leverages XAI techniques for testing DL systems, i.e., systems that use DNNs.\footnote{This paper uses the terms DNNs and DL systems interchangeably, for simplicity of exposition.} In the following, we provide background on foundational concepts used in this paper:~(1)~semantic-based input representation and (2)~XAI methods.

\subsection{Semantic-based Input Representation}
\label{sec:semantic-based-input-representation}

Semantic-based test generation approaches utilize \emph{abstract aspects} of the inputs, such as the shape of a digit in an image or the sentiment polarity of a word in a text, rather than relying on \emph{concrete aspects} of the inputs, such as changes of pixel values in an image or letter modifications in a word~\cite{deepatash,2021-Riccio-ASE,zohdinasab2021deephyperion}. Semantic-based approaches include model-based techniques~\cite{2020-Riccio-FSE}, which utilize domain-specific models to generate test inputs. These methods can translate the input from its concrete representation to its abstract representation and back. By manipulating inputs in this abstract space, perturbations gain semantic meaning, allowing for controlled perturbations that reflect meaningful changes in the original data.
Semantic-based input representation brings an important benefit for test generation--it reduces the dimensionality of the search space.
Conceptually, there exists a 1-to-many mapping from abstract to concrete values, enabling the smaller input space to be more exhaustively covered. Furthermore, semantic validity criteria are expressed at the model level, preventing automated algorithms from creating unrealistic inputs that fall outside the valid operational domain of the DL system~\cite{2023-Riccio-ICSE, jiang2024validity, zhang2024enhancing}.

\subsection{XAI Methods}

XAI methods are used to make the decisions of complex DL systems more transparent and understandable~\cite{vilone2020explainable,gunning2019xai, dovsilovic2018explainable}, which is essential for validation, trust-building, and regulatory compliance.
In this paper, we consider XAI methods and use cases (sentiment analysis, digit recognition, and advanced driving assistance systems or ADAS) established in prior work on DL testing~\cite{2020-Riccio-FSE,2021-Riccio-ASE,deepatash}.

For sentiment analysis, we consider the local explanations produced by \xaiLime~\cite{lime}, \xaiSG~\cite{smilkov2017smoothgrad}, and \xaiIG~\cite{integratedgradients}. In brief, \xaiLime~\cite{lime}, namely local interpretable model-agnostic explanation, explains the predictions of DL systems by approximating their behavior with a simpler, interpretable, surrogate model in a local region. 
However, a large number of perturbations and inferences are required in this process, which makes \xaiLime computationally expensive. Therefore, we also consider other gradient-based XAI methods (\xaiSG and \xaiIG) to ensure the efficiency of test generation. These approaches are representative of different families of XAI methods and address known issues of saliency methods, such as gradient discontinuity and saturation. Particularly, \xaiSG~\cite{smilkov2017smoothgrad} reduces the noise of gradient-based explanations by adding artificial noises and averaging them, whereas \xaiIG~\cite{integratedgradients} overcomes the gradient saturation by summing over scaled inputs.

Handwritten digit recognition and advanced driving assistance are imagery tasks, for which we consider saliency or pixel attribution methods~\cite{simonyan2013deep, shrikumar2017learning,li2021experimental}. These solutions generate local explanations in the input space, identifying regions of an input image that influence the decision-making process of DNNs.
Particularly, in addition to using \xaiSG~\cite{smilkov2017smoothgrad} and \xaiIG~\cite{integratedgradients} mentioned before, we also considered CAM (Class Activation Map)-based methods, as the models under test are typically convolutional neural networks. 
For example, \xaiGC~\cite{Grad-CAM++_2018} generates heatmaps by computing a weighted combination of the positive partial derivatives from the last convolutional layer. 

\section{\tool}\label{sec:approach}

\begin{table}[t]

\footnotesize

\caption{Testing characterization for three popular tasks often solved with DL systems.}
\label{tab:system}

\resizebox{\columnwidth}{!}{%
\begin{tabular}{llll}
    \toprule
    \bf  & \bf Sentiment analysis & \bf Digit recognition & \bf Advanced driving assistance \\
    \midrule
    Testing level & Model & Model & System\\
    DNN task & Classification & Classification  & Regression\\
    Test input & Text & Digit image &  Road in a driving simulator\\
    Input to DL system & Word tokens & Digit image & Driving scenario  \\
    Semantic representation & Words & Control points (image)  & Control points (road center) \\
    Failure criterion & Misclassification & Misclassification & Out of road bounds \\ 
    \bottomrule
\end{tabular}
}
\end{table}

\tool is a focused fuzzing technique for DL systems. It leverages post-hoc local explanations for individual predictions to guide fuzzing. \tool is applicable to any DL system where semantic representations are available. As mentioned in \autoref{sec:semantic-based-input-representation}, we instantiate \tool for tasks established in prior work on DL testing~\cite{2020-Riccio-FSE,2021-Riccio-ASE,deepatash} for which 
semantic models exist: (1)~sentiment analysis, (2)~digit recognition, and (3)~advanced driving assistance systems (ADAS).

\vspace{1ex}\noindent\textbf{Testing scenarios.~}
\autoref{tab:system} shows the characteristics of our testing scenarios. Semantic representations are often domain-specific. For example, both sentiment analysis and digit recognition are classification tasks, but the semantic modelings are different. In sentiment analysis, the words in a movie review are mapped to an embedding space whereas in digit recognition, the bitmap of a MNIST handwritten digit~\cite{mnist} is converted into sequences of cubic Bézier curves~\cite{farin1983algorithms} defined by a series of control points, according to the Scalable Vector Graphics representation. For advanced driving assistance, the centerline of a road in a driving scenario is represented by Catmull-Rom cubic splines~\cite{catmull1974class}, also specified by a sequence of control points. 

The different testing levels also imply distinct interpretations of failures. For sentiment analysis and digit recognition, a failure is defined as the disagreement between the output of the DNN under test and the ground truth, e.g., a misclassification of the movie review or the handwritten digit. For advanced driving assistance, failures are identified by the misbehavior of the whole DL system, i.e., the autonomous vehicle, in response to the outputs of the DNN, rather than isolated incorrect predictions on individual images. Following existing research~\cite{Abdessalem-ICSE18}, we evaluate whether a sequence of incorrect DNN predictions leads to a violation of the safety requirements of the system, i.e., the vehicle driving out of the road, or a reduction of the driving quality~\cite{2021-Jahangirova-ICST}. 

 Notably, unlike evolutionary fuzzing approaches, \tool does not employ a continuous fitness score to rank or retain test inputs. Instead, it solely relies  on a binary failure criterion to decide whether to keep a mutated input. The intention is to isolate and evaluate the impact of XAI-guided mutation alone, without the confounding effects of fitness-based selection. 
 This design also preserves flexibility for integrating \tool into larger testing frameworks that may employ their own prioritization or fitness metrics. 

\begin{algorithm}[t]
    \footnotesize
    \SetAlgoLined
    \KwIn{DNN under test, set of test inputs $T$, failure criterion, termination condition.}
    \KwOut{Failure set $F$}
    
    Initialize test inputs $T$ with random individuals if $T = \emptyset$\;
    Initialize failure set $F = \emptyset$\;
    \While{$T \neq \emptyset$ or termination condition}{
        \For{each test input $t$ in $T$}{
            Test DL system with the input $t$\;
            \eIf{failure criterion not met}{
                $S$ $\leftarrow$ \sc{GetSemanticRepresentation}($t$) \;
                $e$ $\leftarrow$ \sc{LocalExplanationComputation}($t$) \;
                $cs$ $\leftarrow$ \sc{SemanticConceptSelection}($S$, $e$) \;
                $d$ $\leftarrow$ \sc{MutationDirectionComputation}($S$, $e$, $cs$) \;
                $S$ $\leftarrow$ \sc{Mutate}($S$, $cs$, $d$) \;
                $t$ $\leftarrow$ \sc{GenerateConcreteInputFromSemantic}($S$) \;
            }
            {
            Move $t$ from $T$ to $F$\;
            }
        }
    }         
    \Return $F$\;
    \caption{\tool: XAI-guided Fuzzing}
    \label{alg:fuzz}
\end{algorithm}

\vspace{1ex}\noindent\textbf{Algorithm.~}
\autoref{alg:fuzz} shows the main steps of \tool. \tool takes as input the DL system, a set of test inputs $T$, the failure criterion, and a termination condition, such as a time budget or a number of generated inputs. \tool produces as output a set $F$ containing failure-inducing inputs on the DNN under test. The test inputs in $T$ are used to test the DL system. Initially, \tool initializes the set of test inputs $T$, e.g., by randomly selecting seeds from a test dataset, when they are available (for sentiment analysis or digit recognition), or by randomly generating inputs if they are unavailable (for the advanced driving assistance system).
If an original test input $t$ produces a failure, it is discarded. 
The main loop of the algorithm (Lines~7---12) retrieves the semantic representation $S$ (Line~7), which is a sequence of semantic concepts, and the local explanations $e$ (Line~8), which will be used for focused mutations. 
Then, a candidate semantic concept $cs \in S$ is selected for mutation (Line~9), based on the area of highest attention indicated by the local explanation $e$, followed by a mutation of the semantic concept in such area (Lines~10---11).
Finally, the newly generated semantic-based representation is restored to the original input space (Line~12) and used for testing the DL system. In case of failure, the algorithm stores the evolved test input $t$ to the failure set (Lines~14). Finally, the algorithm returns the failure set $F$.

\vspace{-4ex}
\subsection{XAI-guided Mutation Operator} 

\tool's insight is to fuzz test input on the critical areas of the DNN attention to create challenging inputs to the corresponding DL system and, consequently, trigger misbehaviors. \tool's mutation operator selects the semantic concept that is near a high-attention region and, subsequently, performs a targeted mutation in such region. 

\tool's mutation operator comprises three steps: (1)~computing the local explanation (Line~8), (2)~selecting the semantic concept to mutate (Line~9), and (3)~determining the mutation direction and applying mutation (Line~10, 11). 
\underline{First}, \tool uses an XAI technique to compute local explanations associated with a DNN and its test instances~(procedure \textsc{LocalExplanationComputation} on Line~8). Typically, the explanation of the DNN prediction for a given input has the same size as the input, with each dimension of the input space corresponding to an explanation score. 
\underline{Second}, \tool selects the semantic concept to mutate on the high-attention area of the local explanation~(procedure \textsc{SemanticConceptSelection} on Line~9), where an intermediate weight vector ${w}$ is obtained and denotes the importance of semantic concepts with $dim({w}) = dim(S)$.  
For textual inputs, the local explanation corresponds directly to the semantic concept, hence selecting the semantic representation is based on the explanation's magnitude (i.e., ${w} = e$ and $dim(e) = dim({w})$). 
However, for model- or system-level imagery inputs, our approach needs to map the high-dimensional explanations,
onto the lower-dimensional semantic concept space through a function $f$ before the selection, i.e., ${w} = f(e)$ and $dim(e) > dim({w})$. 
To ensure diversity of generated test inputs, \tool\ randomly selects candidate semantic concepts $cs$ based on the weights ${w}$, where semantic concepts with higher weights have higher chances of being chosen.
\underline{Third}, when a candidate semantic concept is selected, \tool leverages the local explanations to choose the mutation direction, aiming to maximize the mutation effectiveness~(Line~10).

\tool applies the same XAI-guided fuzzing workflow across different DL applications, and it is configured according to the kind of input and semantic representation associated with the task under test (\autoref{tab:system}). 
We elaborate on the domain-specific configurations in terms of semantic concept selection and mutation direction in the following sections.

\subsection{Model-level Application on Textual Inputs}\label{sec:2.1 model-level Testing textual}

For textual inputs, the inputs to the DNN are tokens corresponding to the words. Therefore, obtained explanations have the same dimensions as the semantic representation, i.e., words \changed{(\autoref{alg:fuzz}: \textsc{GetSemanticRepresentation})}.

With model-agnostic XAI technique methods such as \xaiLime~\cite{lime}, the feature attributions for each token can be computed directly. 
With gradient-based XAI methods, however, the presence of a non-differentiable embedding layer in the DL model under test, makes the gradient computation unavailable. To address this, we propose a customized approach for interpreting such DL models. Let us assume that the embedding layer mapping the input vector $x$ consisting of $L$ tokens in the embedding space of dimension $N$, is a matrix of size $\mathbb{R}^{L\times N}$.
Then we take the following steps. First, we remove the embedding layer from the model under test to obtain the corresponding submodel. Second, we apply gradient-based XAI methods to compute the explanation $\hat{e} \in \mathbb{R}^{L\times N}$ in the embedding space for the submodel, with each row vector $\hat{e}[i] \in \mathbb{R}^{N}$ representing the explanation vector for the corresponding token $x[i]$. Third, we map the explanation in the embedding space back to the input space with summation (for \xaiIG) or norm operators (for \xaiSG), e.g., $e[i] = sum(\hat{e}[i])$ for each token $x[i]$. 
This process yields the local explanation $e$ with the same dimension as the input $x$, which \tool later uses to guide mutations \changed{(\autoref{alg:fuzz}: \textsc{LocalExplanationComputation})}.

\begin{figure*}[t]
    \centering
    \includegraphics[width=\linewidth]{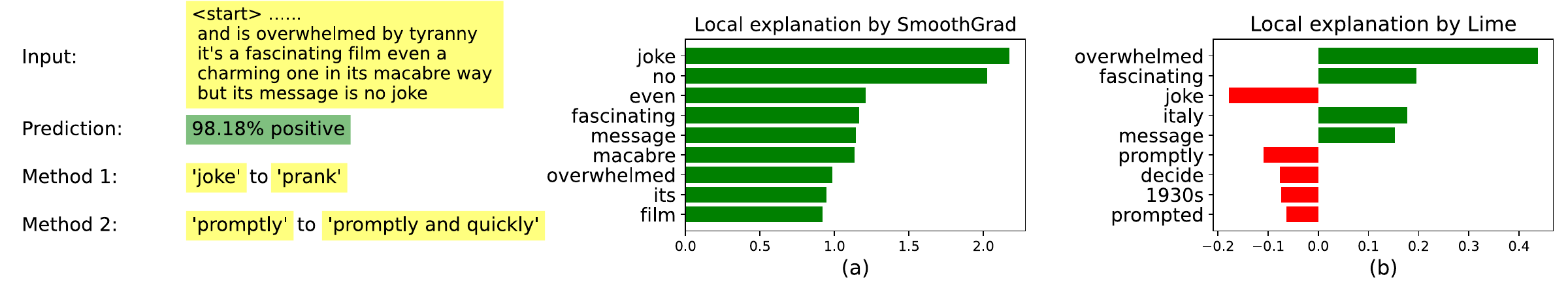}
    \caption{Example of movie review sentiment analysis case study (best viewed in color).
                (a)~Local explanation by \xaiSG;
                (b)~Local explanation by \xaiLime. 
                } 
    \label{fig:imbd method}
\end{figure*}

The explanations are used to derive a weight vector $w$ over tokens, from which \tool samples one candidate concept $cs$ \changed{(\autoref{alg:fuzz}: \textsc{SemanticConceptSelection})}. However, the local explanations obtained for textual inputs differ in their characteristics depending on the XAI method used. 
For instance, \xaiSG captures the gradient information, indicating the sensitivity of the predictions to word variations (\autoref{fig:imbd method} (a)). In this case, the weight vector consists of absolute gradient information. 
In contrast, \xaiLime and \xaiIG provide explanations on feature attributions, 
which represent the magnitude of the positive (or negative) contribution of a word on the prediction, making them useful for guiding the mutation direction (\autoref{fig:imbd method} (b)). In this case, the weight vector contains the raw explanations.

Regarding the actual mutation operation, we used two mutation methods available in the literature~\cite{deepatash}: (1)~replacing a word with its synonym obtained from WordNet~\cite{miller1995wordnet}, and (2)~adding an ``and'' conjunction after an adjective (or adverb), followed by a synonym of the adjective (or adverb). These operators are used to perturb the selected token $cs$, resulting in a modified semantic input $S$ \changed{(\autoref{alg:fuzz}: \textsc{Mutate})}.
These mutation methods ensure that the original meaning of the sentence is preserved, thereby maintaining a high validity of generated test inputs.

During the fuzzing process, one of the mutation methods is selected at each iteration. 
When the first mutation method is chosen, we apply it to the selected word $cs$, regardless of the considered XAI method. This is because it is uncertain whether replacing a synonym will increase or decrease the contribution of the selected word. 
In contrast, the second mutation method enhances the semantics of the sentence by adding a synonym. 
Depending on the explanation, XMutant decides whether to apply directional mutation (e.g., amplifying or attenuating polarity) or random variation \changed{(\autoref{alg:fuzz}: \textsc{MutationDirectionComputation})}.
Therefore, for XAI methods that differentiate feature attributions (i.e., \xaiLime~\cite{lime} and \xaiIG~\cite{integratedgradients}), we perform targeted mutation by applying the second mutation method only to the word with opposite effects to the prediction (e.g., we modify ``promptly'' to ``promptly and quickly'' but we do not modify ``overwhelmed'' in \autoref{fig:imbd method}), which helps to more effectively challenge the model under test. For XAI methods that are not focused on explaining the feature attributions (i.e., \xaiSG), we apply the second mutation method on the selected word, regardless of its effect on the prediction. 

After mutation, the modified semantic input is projected back into the concrete input space—i.e., transformed into a full sentence—which serves as a new test case for the DL system \changed{(\autoref{alg:fuzz}: \textsc{GenerateConcreteInputFromSemantic})}.

\begin{figure*}[t]
\centering
\includegraphics[width=\linewidth]{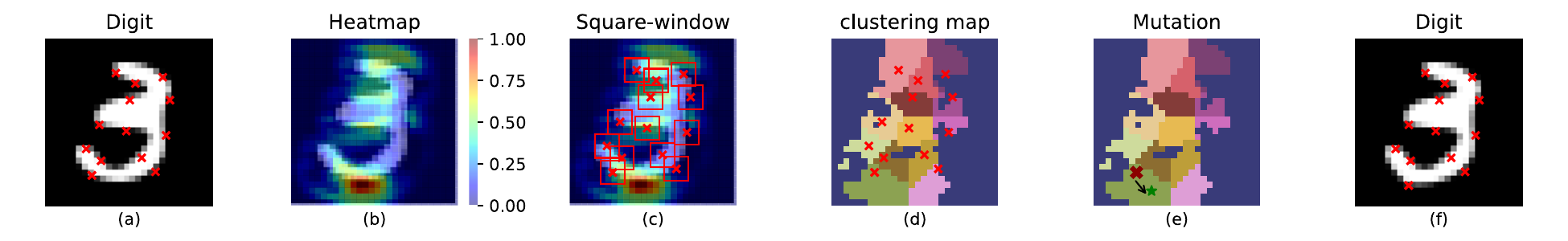}
\caption{\tool steps applied to the MNIST case study (best viewed in color). 
    (a)~SVG semantic-based representation of a digit 3. Control points (i.e., semantic representation) are shown as red crosses;
    (b)~Heatmap generated by \xaiGC; 
    (c)~Square windows centered on the control points;
    (d)~Clustering on the local explanation (one cluster per control point);
    (e)~Control point selection (red cross marker) and mutation towards the attention attractor of the cluster (green star);
    (f)~Digit 3 resulting by mutating the candidate control point. 
    } 
\label{fig:mnist method}
\end{figure*}

\subsection{Model-level Application on Imagery Inputs}\label{sec:2.1 model-level Testing}

For imagery inputs, the semantic representation $S$ is extracted by converting the bitmap image into a set of SVG control points~\changed{(\autoref{alg:fuzz}: \textsc{GetSemanticRepresentation})}. The local explanation, so-called heatmap, is itself an image with the same dimensions as the original image, where pixels are highlighted based on the attention that the DNN under test gives to each portion of the image \changed{(\autoref{alg:fuzz}: \textsc{LocalExplanationComputation)}. 
The subsequent} selection of the candidate semantic concept to be mutated and the computation of the mutation direction are based on a single heatmap $\mathbf{H}$.

\autoref{fig:mnist method} shows the different steps of the \tool operator for a handwritten MNIST~\cite{mnist} digit image. Given an image, \tool converts the bitmap representation into an SVG representation to extract the semantic representation, i.e., a sequence of control points (\autoref{fig:mnist method} (a)). Then, it computes the heatmap on this SVG representation using an XAI method (e.g., \xaiGC in \autoref{fig:mnist method} (b)); in the heatmap relevant locations correspond to hot color intensities (e.g., red/yellow), whereas irrelevant locations correspond to cold color intensities (e.g., blue).
The heatmap values are normalized such that low-attention values are close to zero, while high-attention values are close to one.

Selecting the candidate semantic concept $cs$, i.e., the control point $cp_k$, for mutation requires the computation of the weights corresponding to the control points based on the heatmap (\autoref{alg:fuzz}: \textsc{SemanticConceptSelection}). We evaluate two techniques in this paper, based on \textit{square windows} and \textit{clustering}.

\subsubsection{Square Windows Semantic Concept Selection}

Square windows involve computing the weight of a control point by averaging the intensity values within a square window centered on the control point $(x_k, y_k)$ (e.g., the red squares in \autoref{fig:mnist method} (c)), according to the following equation:

\begin{equation}
     w_k =  \frac{1}{(2d+1)^2} \sum_{i=-d}^{i=d} \sum_{j=-d}^{i=d} \mathbf{H}[(\mathbf{p}_{c_k} + (i, j))],
     \label{eq:sw}
\end{equation}

\noindent
where $d$ is the distance between the center of the square $(x_k, y_k)$ and each side (and the window size $\textit{ws}$ is given by $\textit{ws} = 2d+1$), while $\mathbf{p}_{c_k} \in \mathbb{R}^2$ denotes the coordinates of the control point in the heatmap. Subsequently, the weights ${w}$ for each semantic concept are normalized. 

The square windows approach, although straightforward, involves choosing an appropriate window size $\textit{ws}$. When the window size is large, it can exceed the input's boundaries or have squares of different control points overlap. 
If the window size is small, large regions of the local explanation might be uncovered, potentially missing some high-attention areas. Additionally, this method does not provide any guidance for the direction of mutation, thus the direction is determined randomly (\autoref{alg:fuzz}: \textsc{MutationDirectionComputation}).

\subsubsection{Clustering-based Semantic Concept Selection}

To reduce reliance on hyperparameters and offer better spatial coherence, we also propose a control point selection technique based on clustering.
\tool uses the k-means clustering algorithm~\cite{wu2012advances} to cluster pixels based on their values with the number of clusters $n$ to match the number of control points. 
We modified the typical k-means process by initializing the centroids with the control points and limited the algorithm to a single iteration, without convergence. \autoref{fig:mnist method}~(d) shows 13 clusters identified by different colors, one for each control point. 
Then, \tool computes the weight $w_k$ for each control point $cp_k$ as the sum of the ratios between each pixel value in the cluster of $cp_k$, and the distance between the coordinates of the pixel and the respective control point:

\begin{equation}
     w_k =  \sum_{i=1}^{m_k}\frac{\mathbf{H}[\mathbf{p}_i^k]}{d_i^k}, 
     \text{   with } d_i^k = \max(1, || \mathbf{p}_i^k - \mathbf{p}_{cp_k} || ),
 \label{eq:}
\end{equation}

\noindent
where $m_k$ is the number of pixels in the cluster of control point $cp_k$, $\mathbf{p}_i^k \in \mathbb{R}^2$ denotes the coordinates of $i$-th pixel in the cluster, and $d_i^k$ is the Euclidean distance between $\mathbf{p}_i^k$ and the coordinates of the control point in the heatmap (i.e., $\mathbf{p}_{cp_k}$). To avoid division by zero when $\mathbf{p}_i^k = \mathbf{p}_{cp_k}$, we set the lower bound of $d_i^k$ to one.

Once the weights have been normalized and the control point $cp_k$ sampled (e.g., the control pointing the green cluster in the bottom-left corner of \autoref{fig:mnist method} (e)), \tool mutates the selected control point towards the attention attractor $\mathbf{c}_k$ (\autoref{alg:fuzz}: \textsc{MutationDirectionComputation}), which is defined as the center of intensity of the respective cluster, given by:

\begin{equation}
     \mathbf{c}_k =  \frac{\sum_{i=1}^{m_k}\mathbf{H}[\mathbf{p}^k_i] \mathbf{p}_i^k}{\sum_{i=1}^{m_k}\mathbf{H}[\mathbf{p}^k_i]}.
    \label{eq:2}
\end{equation}
Finally, the new input is obtained by mutating $cs$ toward the attractor $\mathbf{c}_k$ (\autoref{alg:fuzz}: \textsc{Mutate, GenerateConcreteInputFromSemantic}).

\autoref{fig:mnist method} (e) shows an example of how the selected control point in the green cluster (bottom left) is mutated towards the respective attention attractor. This control point is marked by a red cross and moves toward the green star, which represents the attention attractor of the green cluster. The mutation direction is depicted by a black arrow extending from the control point to the attention attractor, leading to a contraction of the high-attention area.
\autoref{fig:mnist method} (f) shows the resulting digit after the mutation, where the control point in the bottom-left corner of the image (see \autoref{fig:mnist method} (a)) has shifted further downward.

\subsection{System-level Application on Logical ADAS Scenarios}\label{sec:2.2System-level Testing}

As the ADAS processes a stream of images from a driving simulator, applying our approach at an individual image level (as described previously for model-level testing) does not effectively assist in selecting and mutating candidate semantic concepts, i.e., control points. Instead, \tool generates a series of heatmaps for each image perceived by the DNN, and aggregates them to retrieve a semantic score that reflects the system's overall performance. This allows the selection of critical semantic components for mutation, following the same fuzzing loop described in \autoref{alg:fuzz}. 

\begin{figure*}[t]
\centering
\includegraphics[width=\linewidth]{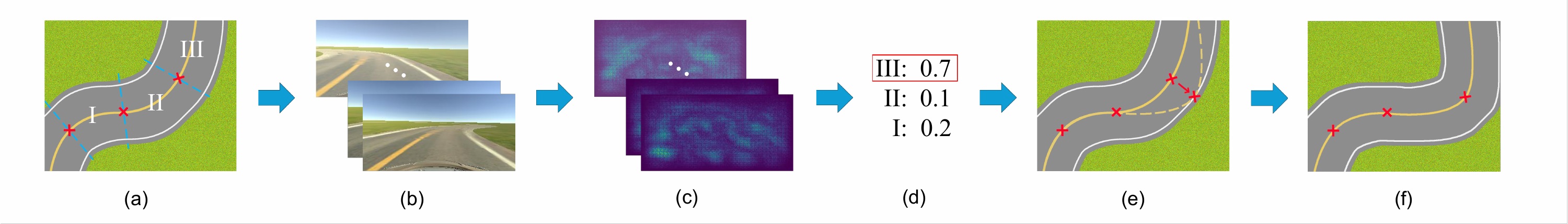}
    \caption{\tool steps for the ADAS (best viewed in color). 
    (a)~A road with 3 control points, or semantic representation (red crosses). The road sectors are indicated as Roman numerals and separated by dashed cyan lines;
    (b)~A sequence of driving frames recorded by the vehicle's camera;
    (c)~A sequence of heatmaps, corresponding to the driving frames, generated by \xaiSG;
    (d)~Weights for each road sector and associated semantic concept;
    (e)~Control point selection (top-right) and mutation towards the right lane (dashed yellow line indicates the new centerline);
    (f)~Road resulting from the mutation of the control point in sector \rom{3}. 
    }
\label{fig:ADS}
\end{figure*}

\autoref{fig:ADS} shows the steps of \tool for the ADAS. The test input is a sequence of control points located at the center of the road (\autoref{fig:ADS} (a)), determining the shape of the two-lane road that will be instantiated in the driving simulator (we assume the other environment variables, such as weather conditions, as fixed). Once the road is instantiated, the DNN that controls the vehicle takes as input the frames that are recorded by the onboard camera (\autoref{fig:ADS} (b)) and outputs the driving commands (i.e., steering angles). Therefore, \tool computes the heatmaps of such camera frames, resulting in a sequence of heatmaps (\autoref{fig:ADS} (c)).

To weigh the control points, we divide the road into segments extending from each control point to the road's centerline (e.g., in \autoref{fig:ADS} (a) there are three road sections). 
Each road segment is initially linked to the control point at its start. This approach is based on preliminary findings indicating that mutations at a control point predominantly affect the driving behavior in the section immediately following it.
To aggregate sequences of heatmaps, \tool computes the derivative of a heatmap  $\nabla \mathbf{H}_t$ at a certain time step $t$ as the average absolute difference between the pixels of two consecutive heatmaps $\mathbf{H}_{t-1}$ and $\mathbf{H}_{t}$:

\begin{equation}
    \nabla \mathbf{H}_t = \frac{1}{wh}\sum_{i=1}^{w}\sum_{j=1}^{h}|\mathbf{H}_{t-1}[i,j] - \mathbf{H}_{t}[i,j]|
\end{equation}

Subsequently, the sampling weight $w_k$ of the control point $cp_k$ is given by the mean value of all derivatives in the corresponding road section, i.e., $w_k = 1/m_k \sum_{i=2}^{m_k} \nabla \mathbf{H}_i^k$, where $m_k$ is the number of frames in a specific road section, and $\{\mathbf{H}_1^k,\dots,\mathbf{H}_{m_k}^k\}$ denotes the heatmaps in the road section associated to the control point $cp_k$.

Indeed, existing work~\cite{2022-Stocco-ASE} has shown that the derivative of a local explanation is indicative of a poor-performing DNN.
In this way, \tool grants a higher weight to road sections---and associated control points---where the driving behavior of the DNN is more erratic \changed{(\autoref{alg:fuzz}: \textsc{SemanticConceptSelection})}. 
Consequently, by mutating such control points in the corresponding road sections, \tool is more likely to generate a test input that induces more challenging driving conditions. 
For instance, in \autoref{fig:ADS} (d), the road section with the highest mean derivative is road section \rom{3}; in this example, the corresponding control point in the top-right corner of \autoref{fig:ADS} (e) is sampled.

Next, \tool determines the mutation direction by considering the perpendicular line to the road's centerline at the candidate control point. Consequently, only two directions are possible: moving the control point toward the right lane or the left lane \changed{(\autoref{alg:fuzz}: \textsc{MutationDirectionComputation})}.
To determine the appropriate mutation direction, we analyze the series of heatmaps for the chosen road segment. 
\tool partitions each local explanation into two equally-sized sub-heatmaps, i.e., $\mathbf{H}[:w/2,:]$ and $\mathbf{H}[w/2+1:,:]$ (where the column separates start and end indices for each dimension of the local explanation).
\tool then calculates the difference in average intensity between the left and right sub-heatmaps in the road segment. The mutation direction is chosen based on the side with the highest average intensity (in \autoref{fig:ADS} (e) the selected direction is the right direction).
The selected control point is then shifted laterally toward that direction \changed{(\autoref{alg:fuzz}: \textsc{Mutate})}, resulting in a new and more challenging road configuration \changed{(\autoref{alg:fuzz}: \textsc{GenerateConcreteInputFromSemantic})}, as visualized in \autoref{fig:ADS} (f).

\section{Empirical Study}\label{sec:study}

\subsection{Research Questions}

We consider the following research questions:

\noindent
\textbf{RQ\textsubscript{1} (effectiveness):} How effective is \tool at finding misbehaviors? 

\noindent
\textbf{RQ\textsubscript{2} (efficiency):}
How efficient is \tool at finding misbehaviors? 

\noindent
\textbf{RQ\textsubscript{3} (configuration):}
How do effectiveness and efficiency vary when considering different XAI algorithms? What is the best configuration of semantic concept selection and mutation direction? 

\noindent
\textbf{RQ\textsubscript{4} (validity):}
To what extent are the inputs generated by \tool valid and label-preserving?

\noindent
\textbf{RQ\textsubscript{5} (comparison):}
How does \tool compare with gradient-guided raw input manipulation techniques?

The first research question (RQ\textsubscript{1}) assesses whether \tool attains a high failure rate.
The second research question (RQ\textsubscript{2}) evaluates the efficiency of \tool.
The third research question (RQ\textsubscript{3}) evaluates different configurations of \tool obtained by varying the XAI algorithm, semantic concept selection, and mutation direction strategy.
The fourth research question (RQ\textsubscript{4}) studies the usefulness of the inputs that \tool produces, both for automated input validators and human assessors.
The final research question (RQ\textsubscript{5}) examines how \tool compares to widely used gradient-guided raw input manipulation techniques. Although these techniques differ in their primary objectives, generalization and robustness testing are often confused since they share the same goal, i.e., DNN reliability. Thus, we analyze and compare the generated inputs from our approach and those generated by existing methods to provide evidence of the benefits of a semantic-based representation, as test inputs generated by these distinct methodologies may form different manifolds in the semantic space.
  
\subsection{Objects of Study} 

\subsubsection{Datasets and Models} 

\textbf{IMDB.} Concerning sentiment analysis, we consider a DL system designed to classify the sentiment (i.e., positive or negative) of movie reviews from the IMDB database~\cite{imbd}. We use the DL system available in the replication package of existing work~\cite{man_sentimental_imdb_lstm}, characterized by an embedding layer, an LSTM layer~\cite{hochreiter1997long}, and two fully connected layers. The model achieves 85.25\% accuracy on the test set. \\
\textbf{MNIST.}
Concerning digit recognition, the DL system classifies handwritten digits from the MNIST dataset~\cite{lecun1998gradient}. This DL system takes 28x28 greyscale images as input and predicts the corresponding digit (the possible classes range from 0 to 9). 
In this paper, we test the convolutional DNN architecture provided in the replication package of existing work~\cite{2020-Riccio-FSE}. 
Architecturally, it is characterized by 2 convolutional layer, a max-pooling layer, and two fully connected layers. 
The model achieves 98.99\% accuracy on the test set.\\
\textbf{ADAS.}
Concerning advanced driving assistance, the DL system controls a vehicle in the Udacity simulator~\cite{udacity-simulator}, a cross-platform driving simulator developed with Unity3D~\cite{unity}, widely used in the ADS testing literature~\cite{2020-Stocco-ICSE,2020-Riccio-EMSE,2021-Jahangirova-ICST,10.1007/s10515-021-00310-0}. The DL system includes a \davetwo model (a DNN regressor), a Level~2~\cite{iso26262} ADAS that performs the lane keeping functionality from a training set of images collected when the driver is an expert human pilot, by predicting the corresponding driving commands imitating the human driving behavior. 
We obtained the trained \davetwo model and the simulator from the replication package of existing work~\cite{2024-Biagiola-EMSE}. 
The model architecture includes three convolutional layers for feature extraction, followed by five fully connected layers. 
The simulator supports the creation of open-loop road tracks for testing ADAS models, including the ability to generate and modify road topologies.

\subsubsection{Comparison Baselines} 

To assess the guidance provided by our XAI-based approach, we compare \tool against \deepjanus, a popular semantic-based test generator~\cite{2020-Riccio-FSE}.  
For the selection of semantic concepts and mutation directions, \deepjanus relies on a random method that selects a candidate semantic concept for mutation based on a uniformly weighted probability. The mutation direction is task-specific. 
For IMDB, it applies one of the mutation methods on a randomly selected semantic concept.
For \mnist, the mutation direction is sampled from the uniform distribution over $0$ to $2\pi$. 
For ADAS, the mutation direction is randomly selected from one of the two directions perpendicular to the road's curvature. 
To ensure a fair comparison, we applied the same extent value for mutation direction to both \tool and \deepjanus's mutation methods. In our study, we sampled the extent of MNIST from a uniform distribution over the interval $[0, 1.2]$, and for ADAS we adopted a fixed value 4, proportional to the road's width 8. These values are determined based on existing literature~\cite{2020-Riccio-FSE,2024-Biagiola-EMSE}.

Regarding raw input manipulation approaches, we selected popular approaches that use gradients to guide the generation of adversarial inputs, i.e., FGSM~\cite{Goodfellow2014ExplainingAH}, DeepXplore~\cite{pei2017deepxplore}, and DLFuzz~\cite{guo2018dlfuzz}.
For FGSM, we progressively increased the perturbation intensity up to the 20\% threshold, as higher intensities affect the validity of the generated inputs (i.e., the original images were no longer recognizable due to excessive noise or corruption). 
For DLFuzz and DeepXplore, we adopted the default settings described in the original papers.

\subsubsection{Metrics used for Analysis} 

Concerning RQ\textsubscript{1}, we evaluate the effectiveness of the test generation technique by computing the \textit{cumulative failure rate} observed under a given number of mutation iterations across different configurations. The cumulative failure rate is obtained by dividing the number of failures by the total number of inputs. For IMDB and \mnist, a failure is characterized by the number of misclassifications. For ADAS, we measure the number of safety-critical failures in terms of \textit{out of bounds} (OOBs), which occur when the vehicle drives outside the road's drivable lanes during execution in the simulator. 

Concerning RQ\textsubscript{2}, we evaluate the performance of \tool by measuring the \textit{relative efficiency} with respect to \deepjanus. Particularly, relative efficiency is computed by integrating the \textit{cumulative failure rates} over the number of iterations to obtain the area under the cumulative failure curve (AUFC). For each configuration and number of iterations, we calculate the AUFC (Area Under the Failure Curve) and then divide it by the AUFC of our comparison baseline. Considering computational time complexity, \tool introduces an additional overhead for computing local explanations compared to the baseline, while all other components in the fuzzing loop introduce the same computational expense. To assess its impact, we specifically evaluated the proportion of the XAI overhead relative to the baseline's average computation time per iteration.
To measure computational cost, we compute the \textit{XAI overhead}, defined as the average time per iteration spent on local explanation, normalized by the baseline (\deepjanus) iteration time:

\[
\text{XAI overhead} = \frac{T_{\tool} - T_{\text{baseline}}}{T_{\text{baseline}}}
\]

To jointly assess these two aspects, relative efficiency and computational cost, we also report a composite efficiency metric, which adjusts the relative AUFC by factoring in the per-iteration overhead, i.e.,
\[
\text{composite efficiency} = \frac{\text{relative efficiency}}{1 + \text{XAI overhead}}
\]
offering a more operational view on real-world efficiency under time constraints.

Concerning RQ\textsubscript{3}, we assess the effectiveness and efficiency of each configuration of \tool using the metrics used for RQ\textsubscript{1} and RQ\textsubscript{2}.
In RQ\textsubscript{4}, we present the validity rates for both automated input validation and human evaluation, along with the label-preservation rates as judged by human assessors. 

In RQ\textsubscript{5}, we evaluate efficiency and effectiveness of competing approaches, by calculating the number of misclassified inputs divided by the total number of original seeds, and the total elapsed time divided by the number of misclassified inputs. 
We also adopt two metrics to evaluate the realism of generated inputs. The density and coverage metrics~\cite{ferjad2020icml} measure how well the generated inputs align with real data distribution, using k-nearest neighbors to define local manifolds. Particularly, coverage quantifies the proportion of real points with nearby generated points, while density reflects the concentration of generated points around the original test inputs. Following the reference implementation in Dola et al.~\cite{dola2024cit4dnn} we set $k=5$.

\subsubsection{Configurations} 

We evaluate a total of 24 configurations, 
designed to systematically assess the impact of the core components (XAI algorithm, the semantic concept selection method, and the mutation direction) of \tool for the purpose of ablation-style analysis.

For IMDB, \tool uses three XAI algorithms (\xaiSG~\cite{smilkov2017smoothgrad}, \xaiLime~\cite{lime}, and \xaiIG~\cite{integratedgradients}) and two mutation methods from the literature, namely synonym replacement and addition~\cite{zohdinasab2021deephyperion}.

In the MNIST classification task, \tool uses three XAI algorithms (\xaiSG~\cite{smilkov2017smoothgrad}, \xaiGC~\cite{Grad-CAM++_2018}, and \xaiIG~\cite{integratedgradients}) and two strategies (square windows and clustering). For square windows, we use a window size value of $\textit{ws}=3$, which was found appropriate during preliminary experiments
(e.g., such value does not lead to the creation of square windows that exceed the input boundaries), as presented in Appendix~10.1.
For mutation direction, \tool relies on random selection due to the absence of attractor information.
The clustering strategy allows for the mutation direction to be chosen either towards the attention attractor or randomly, thereby isolating the effect of mutation direction. Additionally, following previous works~\cite{fongInterpretableExplanations,selvaraju2017grad}, \tool pre-processes the heatmap to discard low-intensity values and filter out the noise using a threshold $\epsilon = 0.1$, improving stability of semantic selection.

For the ADAS regression task, \tool uses three XAI algorithms (\xaiSG~\cite{smilkov2017smoothgrad}, \xaiGC~\cite{Grad-CAM++_2018}, and \xaiIG~\cite{integratedgradients}). Following existing literature, it adopts the heatmap derivative function method from ThirdEye~\cite{2022-Stocco-ASE} to aggregate consecutive heatmaps into a single score.
Considering the mutation direction, \tool uses three strategies (the attractor mutation direction does not apply to ADAS).
The first strategy determines mutation directions by identifying the lane receiving the most focus (High). For instance, if the heatmaps indicate higher attention on the right lane, the mutation direction chosen is to the right. In contrast, the second strategy involves mutation directions in lanes receiving the least focus (Low). 
For example, if the heatmaps reveal the right lane as receiving the most focus, the mutation direction is set to the left. The final strategy adopts a random approach to select the mutation direction (i.e., either left or right).

These controlled variations across tasks and configurations enable us to analyze the contribution of each component in guiding semantic-based fuzzing.

\subsubsection{Procedure}

Concerning RQ\textsubscript{1}, RQ\textsubscript{2}, and RQ\textsubscript{3}, the evaluation procedure is as follows. 

For IMDB, we randomly select $200$ movie reviews from the original \imdb test set. We discard $26$ reviews that are misclassified before mutation.
For \mnist, we randomly select 2,000 digits from the original \mnist\ test set, uniformly distributed for each class. We run the classifier to ensure that all original seeds are correctly classified. We discard $23$ digits identified during this screening task and evaluate the remaining seeds.
For ADAS, we randomly generate $60$ test inputs (i.e., roads) with the following characteristics: maximum curvature $70$ degrees and $12$ control points to avoid too many invalid roads and ensure the variety of choice during mutation~\cite{2024-Biagiola-EMSE}. 
As a sanity check, we ensured that the trained \davetwo model drives all roads in the simulator successfully. As a result, we discard $5$ roads that trigger misbehaviors before any mutations.
We execute each configuration of \tool and \deepjanus, using a budget of $100$ iterations for IMDB, 1,000 for \mnist, and $30$ for the ADAS. This upper bound value was found through experimentation to be adequate for convergence of either \tool or \deepjanus.

Concerning RQ\textsubscript{4}, we assess the validity of \tool's output using automated validators and questionnaires for human assessment. Particularly, for IMDB, we perform the validity check using ChatGPT~\cite{openai}, a state-of-the-art large language model. For textual inputs, manual validation is inherently subjective and can lack consistency. Additionally, the workload for manual evaluation is significant, potentially leading to decreased evaluation quality. 
Large language models like ChatGPT, on the other hand, provide a viable automated solution for validation and are increasingly regarded as reliable as humans for comprehending and explaining small chunks of text, such as those produced by XMutant for IMDB, as highlighted by recent studies~\cite{wang2025llmsreplacehumanevaluators,zheng2023judgingllmasajudgemtbenchchatbot}.

We used ChatGPT-4o-mini~\cite{GPT} since it is cost-efficient and shows strong performance on textual intelligence and reasoning tasks.
We append the textual input (to be classified) to the following prompt ``\textit{Assume you are a sentiment classifier, given a text of a movie review removing Stopwords and Punctuations. Please only reply `positive', `negative', or `invalid' if the sentence does not make sense.}''.
Due to the inherent randomness of ChatGPT, we repeated the validity assessment five times for each input and reported the average score.

For \mnist, we randomly select 200 misclassified inputs produced by the best configuration of \tool and the \deepjanus baseline. We assess the validity of \tool's and \deepjanus's output using SelfOracle~\cite{2020-Stocco-ICSE}, a distribution-aware input validator for imagery data, which has shown high agreement with human validity assessment in a large comparative study about test input generators for DL~\cite{2023-Riccio-ICSE}. We obtained the trained model of SelfOracle for MNIST from the replication package of the paper by Riccio and Tonella~\cite{2023-Riccio-ICSE}. We applied SelfOracle to reconstruct all failure-inducing images by each variant of \tool and \deepjanus using the same rate of false alarms as the original study ($\epsilon=0.05$\%).
We also evaluate validity and label preservation with human assessors. We conducted a questionnaire where 10 participants were asked to identify a digit, with choices ranging from 0 to 9, or to indicate if it was unrecognizable as a digit. This method allowed us to assess both the validity (if the response is a digit) and label preservation (if the response matches the intended label).

For ADAS, we did not perform any automated or manual validation of the roads, as our test generation process filters out invalid roads by ensuring they meet specific criteria: (1)~the start and end points are different; (2)~the road is contained within a square bounding box of a predefined size (specifically 250 $\times$ 250), and (3)~there are no intersections.

Concerning RQ\textsubscript{5}, we conduct the comparison only for the image classification task, as the comparing techniques do not apply to textual inputs and logical driving scenarios. 
We use the same original 2,000 seeds for all techniques, for which we compute the effectiveness and efficiency metrics. 
For coverage analysis, we randomly selected 50 generated failure-inducing inputs per class, totaling 500, to ensure that coverage-related metrics are not affected by various population sizes. However, for DeepXplore, we only obtain 44 generated inputs out of 2,000 original seeds, so we analyze all available samples.

Since our analysis prioritizes semantic similarity instead of pixel-wise similarity, we conduct coverage analysis in an embedding space that captures high-level semantic features. We use a pretrained VGG16 model on ImageNet as an embedding extractor, as it has demonstrated 100\% accuracy in MNIST classification through transfer learning \cite{kaggle_vgg16_mnist}, confirming its ability to capture relevant digits features. 
The 512-dimensional embedding vectors are then visualized by reducing their dimensionality using principal component analysis (PCA), to show the overlap between generated and real data distributions.

\subsection{Results}

\changed{\subsubsection{Effectiveness (RQ\textsubscript{1})}}

Concerning effectiveness (RQ\textsubscript{1}), \autoref{tab:rq1} presents the effectiveness result (cumulative failure rate) for all configurations of \tool and \deepjanus as the baseline, for all case studies (IMDB, MNIST, and ADAS), over different iterations.
For all case studies, all configurations of \tool outperform the baseline, regardless of the iteration considered. 
We assessed the statistical significance of the differences in the cumulative failure rate between \tool and the baseline using the non-parametric Mann-Whitney U test~\cite{Wilcoxon1945} (with $\alpha = 0.01$) and the magnitude of the differences using the Cohen's $d$ effect size~\cite{cohen1988statistical}. 
The differences were found to be statistically significant ($p$-value $<$ 0.01), with varying effect sizes. 


\begin{table*}[t]
\renewcommand{\arraystretch}{1}
\setlength{\tabcolsep}{1.5pt} 

\centering

\scriptsize

\caption{\changed{Results for Effectiveness (RQ\textsubscript{1}). Values indicate the \textit{Cumulative Failure Rate [\%]} over different Iterations. P-values with statistically significant differences are boldfaced.}}

\resizebox{\columnwidth}{!}{%

\begin{tabular}{c@{\hspace{20pt}}c@{\hspace{36pt}}c@{\hspace{46pt}}c@{\hspace{45pt}}c@{\hspace{45pt}}c}

\toprule

&Iterations & \xaiSG & \xaiLime & \xaiIG & DJ \\
\midrule
& 20	& 21.84	& 32.18 & 31.03 & 9.77\\
& 40	& 39.08 & 50.57 & 56.90 & 17.82 \\
& 60	& 46.55 & 62.64 & 66.67 & 24.14\\
& 80	& 50.57 & 69.54 & 73.56 & 28.74\\
\multirow{-5}{*}{\rotatebox[origin=c]{90}{IMDB}} & 100   & 54.60	& 76.44	& 75.86	& 32.18\\

\midrule
						
& { p-value } &\bf 6.93E-18  & \bf 1.17E-17 &\bf 7.90E-18 & -\\
\multirow{-2}{*}{} & {effect size} & large & large & large & -\\


\end{tabular}
}

\resizebox{\textwidth}{!}{
\begin{tabular}{cccccccccccc}

\toprule

\multicolumn{2}{c}{} & \multicolumn{3}{c}{\xaiSG} & \multicolumn{3}{c}{\xaiGC} 
& \multicolumn{3}{c}{\xaiIG} & {DJ} \\
\cmidrule(r){3-5}
\cmidrule(r){6-8}
\cmidrule(r){9-11}

\multicolumn{2}{c}{} & \multicolumn{2}{c}{Clustering} & \multicolumn{1}{c}{SW} & \multicolumn{2}{c}{Clustering} & \multicolumn{1}{c}{SW} 
& \multicolumn{2}{c}{Clustering} & \multicolumn{1}{c}{SW} & - \\

\cmidrule(r){3-4}
\cmidrule(r){5-5}
\cmidrule(r){6-7}
\cmidrule(r){8-8}
\cmidrule(r){9-10}
\cmidrule(r){11-11}

\multicolumn{2}{c}{{Iterations}} & Attractor & {Random} & {Random} & Attractor & {Random} & {Random} 
& Attractor & {Random} & {Random} & - \\ 

\midrule

& {200}  & 77 & 24 & 24 & 74 & 21 & 22 & 60 & 27 & 26 & 15 \\
& {400}  & 94 & 47 & 45 & 94 & 46 & 45 & 80 & 51 & 47 & 35 \\
& {600}  & 97 & 62 & 60 & 98 & 60 & 61 & 87 & 66 & 61 & 51 \\
& {800}  & 98 & 71 & 69 & 99 & 70 & 73 & 90 & 76 & 71 & 62 \\
\multirow{-5}{*}{\rotatebox[origin=c]{90}{MNIST}} & {1000} & 99 & 78 & 76 & 99 & 77 & 80 & 92 & 82 & 77 & 70 \\

\midrule
						
& {p-value } & {\bf  4.94E-165} & {\bf 1.09E-164} & {\bf 3.46E-165} & {\bf 2.57E-164} & {\bf 1.19E-161} &{ \bf 2.10E-164} & {\bf 4.86E-165} & {\bf 7.03E-165} &{\bf  7.05E-165} & - \\ 

\multirow{-2}{*}{} & {effect size} & {large} & {small} & {small} & {large} & {small} & {small} & {large} & {medium} &{small} & - \\

\toprule

\multicolumn{2}{c}{} & \multicolumn{3}{c}{\xaiSG} & \multicolumn{3}{c}{\xaiGC}  & \multicolumn{3}{c}{\xaiIG} & {DJ} \\

\cmidrule(r){3-5}
\cmidrule(r){6-8}
\cmidrule(r){9-11}

\multicolumn{2}{c}{Iterations} & {Random} & {Low} & {High} & {Random} & {Low} & {High} & {Random} & {Low} & {High} & {-} \\ \midrule
 & {{ 10}}  & 65 & 60 & 67 & 75 & 87 & 75 & 64 & 65 & 58 & 51 \\
 & {{ 20}}  & 89 & 82 & 84 & 95 & 93 & 95 & 87 & 85 & 85 & 75 \\
& {{ 30}}  & 91 & 89 & 95 & 95 & 95 & 95 & 95 & 87 & 89 & 84 \\
\multirow{-4}{*}{\rotatebox[origin=c]{90}{ADAS}} & {{ 40}}  & 95 & 89 & 95 & 95 & 95 & 95 & 95 & 93 & 95 & 95 \\

\midrule

&{p-value} & {\bf 8.23E-08} & {\bf 5.26E-06 } & {\bf 1.05E-07} & {\bf 1.07E-07} & {\bf 1.10E-07} &{\bf  1.09E-07} & {\bf 1.05E-07} & {\bf 1.54E-07} &{\bf 1.07E-07} & -\\
\multirow{-2}{*}{} & {effect size} & {small} & {small} & {small} & {medium} & {medium} &{medium} & {small} & {small} &{small} & - \\

\bottomrule

\end{tabular}
} 
\label{tab:rq1}
\end{table*}


\begin{table*}[t]
\renewcommand{\arraystretch}{1}
\setlength{\tabcolsep}{1.5pt}

\centering

\scriptsize

\caption{\changed{Results for Efficiency (RQ\textsubscript{2}). Values indicate the \textit{Relative Efficiency} (AUFC Ratio) over different Iterations, followed by XAI Computation Overhead and composite time-to-failure efficiency compared to the \deepjanus. The Iteration used for the computation of composite efficiency is underlined. The composite efficiency values outperforming baseline are boldfaced.}}

\resizebox{\columnwidth}{!}{%

\begin{tabular}{c@{\hspace{20pt}}c@{\hspace{36pt}}c@{\hspace{46pt}}c@{\hspace{45pt}}c@{\hspace{45pt}}c}

\toprule

&Iterations & \xaiSG & \xaiLime & \xaiIG & DJ \\ 
\midrule
& 10 & 1.93 & 1.44 & 2.05 & 1 \\
& 20 & 2.08 & 2.31 & 2.71 & 1 \\
& \underline {40} &\underline {2.12} &\underline{2.82} &\underline {3.05} & 1 \\
\multirow{-4}{*}{\rotatebox[origin=c]{90}{IMDB}} & 100 & 1.90 & 2.74 & 2.59 & 1 \\

\midrule

& { XAI overhead $\downarrow$} & 0.35  & 236.63 & 1.47 & 0\\
&{Composite Efficiency $\uparrow$} &\textbf{ 1.57} & 0.01 & \textbf{1.23} & 1\\

\end{tabular}
}

\resizebox{\textwidth}{!}{
\begin{tabular}{cccccccccccc}

\toprule

\multicolumn{2}{c}{} & \multicolumn{3}{c}{\xaiSG} & \multicolumn{3}{c}{\xaiGC} 
& \multicolumn{3}{c}{\xaiIG} & {DJ} \\
\cmidrule(r){3-5}
\cmidrule(r){6-8}
\cmidrule(r){9-11}

\multicolumn{2}{c}{} & \multicolumn{2}{c}{Clustering} & \multicolumn{1}{c}{SW} & \multicolumn{2}{c}{Clustering} & \multicolumn{1}{c}{SW} 
& \multicolumn{2}{c}{Clustering} & \multicolumn{1}{c}{SW} & - \\

\cmidrule(r){3-4}
\cmidrule(r){5-5}
\cmidrule(r){6-7}
\cmidrule(r){8-8}
\cmidrule(r){9-10}
\cmidrule(r){11-11}

\multicolumn{2}{c}{{Iterations}} & Attractor & {Random} & {Random} & Attractor & {Random} & {Random} 
& Attractor & {Random} & {Random} & - \\ 

\midrule

&\underline {100}  & \underline {7.00} &\underline  {1.83} & \underline {2.11} &\underline { 4.81} & \underline {1.03} &\underline  {1.32} & \underline {5.54} & \underline {2.06} &\underline { 2.15} & 1 \\
&200  & 6.37 & 1.77 & 1.82 & 5.54 & 1.32 & 1.48 & 4.92 & 2.00 & 1.97 & 1 \\
&400  & 4.03 & 1.52 & 1.47 & 3.84 & 1.34 & 1.36 & 3.26 & 1.67 & 1.57 & 1 \\
\multirow{-4}{*}{\rotatebox[origin=c]{90}{MNIST}} &1000 & 2.12 & 1.23 & 1.20 & 2.09 & 1.19 & 1.21 & 1.86 & 1.32 & 1.23 & 1 \\

\midrule
						
& { XAI overhead $\downarrow$} & { 0.87} & {0.87} & {0.87} & {0.79} & {0.79} &{ 0.79} & {0.52} & {0.52} &{0.52} & 0 \\ 
& {Composite Efficiency $\uparrow$} & {\textbf{3.74}} & {0.98} & {\textbf{1.13}} & {\textbf{2.69}} & {0.58} & {0.74} & {\textbf{3.64}} & {\textbf{1.36}} &{\textbf{1.41}} & 1 \\

\toprule

\multicolumn{2}{c}{} & \multicolumn{3}{c}{\xaiSG} & \multicolumn{3}{c}{\xaiGC}  & \multicolumn{3}{c}{\xaiIG} & {DJ} \\

\cmidrule(r){3-5}
\cmidrule(r){6-8}
\cmidrule(r){9-11}

\multicolumn{2}{c}{Iterations} & {Random} & {Low} & {High} & {Random} & {Low} & {High} & {Random} & {Low} & {High} & {-} \\ \midrule
&\underline{5}  & \underline{2.00} &\underline{ 2.20 }&\underline {2.87} &\underline{ 2.80} &\underline {2.53} &\underline {2.07} &\underline {2.13} &\underline{ 2.53} &\underline {2.60} & 1 \\
&10 & 1.35 & 1.36 & 1.69 & 1.86 & 1.78 & 1.60 & 1.57 & 1.60 & 1.56 & 1 \\
&20 & 1.31 & 1.21 & 1.36 & 1.47 & 1.53 & 1.48 & 1.30 & 1.30 & 1.32 & 1 \\
\multirow{-4}{*}{\rotatebox[origin=c]{90}{ADAS}} &40 & 1.15 & 1.09 & 1.18 & 1.23 & 1.25 & 1.23 & 1.16 & 1.12 & 1.14 & 1 \\

\midrule

&{ XAI overhead $\downarrow$} & {0.19} & 0.19 & { 0.19} & { 0.16} & {0.16} &{0.16 } & { 0.16} & {0.16} &{0.16 } & 0\\

& {Composite Efficiency $\uparrow$} & \textbf{1.68} & \textbf{1.85} & \textbf{2.41} & \textbf{2.41} & \textbf{2.18} &\textbf{1.78} & \textbf{1.83} & \textbf{2.18} &\textbf{2.24} & 1 \\

\bottomrule

\end{tabular}
} 
\label{tab:rq2}
\end{table*}

\begin{tcolorbox}[boxrule=0pt,sharp corners,boxsep=2pt,left=2pt,right=2pt,top=2.5pt,bottom=2pt]
\begin{center}
\begin{minipage}[t]{0.99\linewidth}
\textbf{RQ\textsubscript{1}}: \textit{
The guidance provided by the XAI allows \tool to 
generate significantly more failure-inducing inputs than \deepjanus (up to $+208\%$ for sentiment analysis, up to $+125\%$ for digit recognition, and $+27\%$ for system-level advanced driving assistance).
}
\label{RQ1}
\end{minipage}
\end{center}
\end{tcolorbox}

\changed{\subsubsection{Efficiency (RQ\textsubscript{2})}}
Concerning performance (RQ\textsubscript{2}), 
\autoref{tab:rq2} presents (1) the relative efficiency, i.e., the ability to induce failures quickly, measured by the relative AUFC w.r.t. \deepjanus; (2) the XAI overhead, i.e., the additional per-iteration cost introduced by XAI computations as a ratio of the baseline's average computation time per iteration; and the joint metric composite efficiency, which reflects the real-world time effectiveness by balancing AUFC gains and computational costs.
Regardless of the XAI overhead, all configurations of \tool were producing misbehaviors faster than the baseline, as evidenced by the relative efficiencies regarding \deepjanus as the relative efficiencies are greater than 1. The results hold for all case studies. 
Considering the XAI overhead, \xaiLime is extremely time-consuming since it requires estimating a surrogate model, making it unsuitable for test generation. 
17 out of 21 configurations of \tool with other XAI methods remains more efficient than \deepjanus, even when the overhead is accounted for, reported by the composite efficiencies larger than 1.
Besides, in system-level ADAS testing, XAI computations constitute a relatively small portion of the total execution time, as executing the driving simulation is computationally expensive.
These results show \tool's time-efficiency, especially for complex system-level testing settings.

\begin{tcolorbox}[boxrule=0pt,sharp corners,boxsep=2pt,left=2pt,right=2pt,top=2.5pt,bottom=2pt]
\begin{center}
\begin{minipage}[t]{0.99\linewidth}
\textbf{RQ\textsubscript{2}}: \textit{
\tool is faster than \deepjanus at exposing failure-inducing inputs for DL systems (ranging from 2$\times$ to 7$\times$ times faster for model-level sentiment analysis and digit recognition, and 2$\times$ faster for system-level advanced driving assistance).
}
\end{minipage}
\end{center}
\end{tcolorbox}

\changed{\subsubsection{Configuration (RQ\textsubscript{3})}} 
Concerning the configurations of \tool (RQ\textsubscript{3}), 
in the case of IMDB, when \tool is equipped with XAI methods that explain the feature attribution (\xaiIG and \xaiLime),
it outperforms the baseline both in terms of cumulative failure rate and relative efficiency. However, since the computation time by \xaiLime has shown to be computationally expensive, the configuration using \xaiIG is more practical for real-world scenarios.

For MNIST, the performance of \tool shows negligible differences in both cumulative failure rate and relative efficiency, regardless of the XAI method used, when relying on different control point selection strategies such as Clustering or Square Windows. These configurations consistently achieved a cumulative failure rate over $75\%$ after 1,000 iterations. Notably, when equipped with clustering-based control point selection and random mutation direction, \xaiIG proved to be the best configuration.
However, the introduction of a mutation direction towards the attractors of the heatmaps significantly enhances \tool effectiveness. All XAI methods, aside from \xaiIG, reached a failure rate of 70\% in less than 200 iterations. This performance is on par with that achieved by \deepjanus over 1,000 iterations. 

Concerning the ADAS task, the relative efficiency of various configurations with \tool is approximately twice that of the baseline. However, combining with heatmap-guided mutation direction did not significantly enhance \tool. Among four different XAI methods, GradCAM++ performed best, achieving a $95\%$ cumulative failure rate in only 15 iterations.

\begin{tcolorbox}[boxrule=0pt,sharp corners,boxsep=2pt,left=2pt,right=2pt,top=2.5pt,bottom=2pt]
\begin{center}
\begin{minipage}[t]{0.99\linewidth}
\textbf{RQ\textsubscript{3}}: \textit{
For model-level digit recognition, the best configurations of \tool are \xaiSG, and \xaiGC equipped with clustering selection and mutation towards the Attractor. For system-level advanced driving assistance, the best configurations of \tool are \xaiGC with the Random mutation or towards the High attention lane.
}
\end{minipage}
\end{center}
\end{tcolorbox}

\subsubsection{Validity (RQ\textsubscript{4})}

\begin{figure}[t]
  \centering

  \begin{subfigure}[b]{0.49\linewidth}
    \centering
    \includegraphics[width=\linewidth]{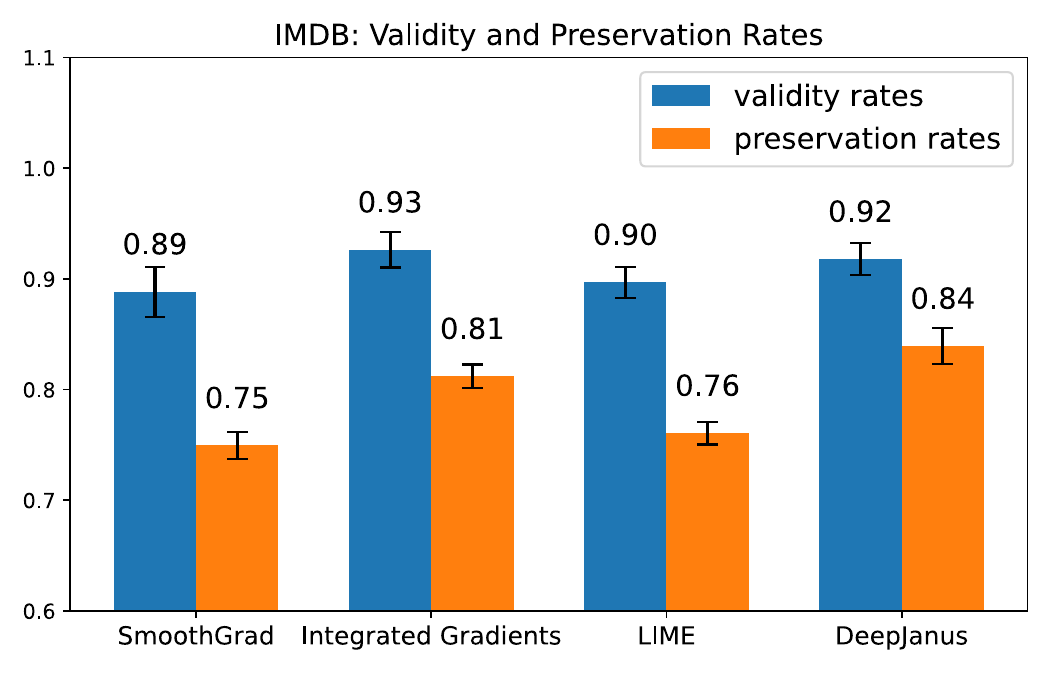} 
    \caption{\imdb.}
    \label{fig:validity_imdb}
  \end{subfigure}
  \hfill
  \begin{subfigure}[b]{0.49\linewidth}
    \centering
    \includegraphics[width=\linewidth]{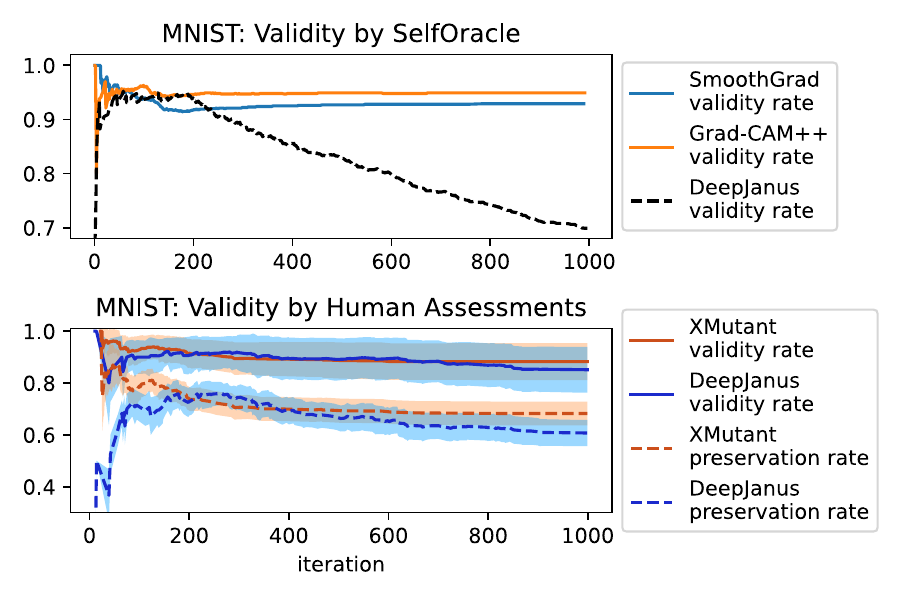}
    \caption{\mnist.}
    \label{fig:validity_mnist}
  \end{subfigure}
  \caption{Validity and preservation rates for sentiment analysis (left) and digit recognition (right).
  }
   \label{fig:validity}
\end{figure}

\autoref{fig:validity} reports the validity rate for \imdb and \mnist. 
\autoref{fig:validity} (a) presents the validity assessment by ChatGPT, including the mean and variance due to ChatGPT’s inherent randomness. All configurations of \tool and \deepjanus maintain a high validity (above 89\%) and preservation rate (above 75\%) with relatively low deviations since the adopted mutation methods are considered conservative and can preserve the input semantics. 
\deepjanus has a slightly higher preservation rate as it was assessed on a smaller set of inputs (\autoref{RQ1}).

For the automated validator of \mnist, we only report the best configurations found in the previous research questions, namely \xaiSG Clustering Attractor and \xaiGC Clustering Attractor. 
The plot (\autoref{fig:validity} (b) top) shows that the validity rate for the \deepjanus steeply decreases with the increasing number of iterations. This is expected because \deepjanus produces out-of-distribution inputs if random mutations are applied an excessive number of times (particularly after 200 iterations). Differently, all configurations of \tool produce failure-inducing inputs that are regarded as in-distribution for SelfOracle. 
It is important to notice that the validity rate for \tool stabilizes at $95\%$ (the used validity threshold in SelfOracle), which indicates that the distribution of the misclassified inputs generated by \tool highly overlaps with the distribution of the original testing dataset of MNIST. This is attributed to the \tool's ability to generate failure-inducing inputs by modifying the original digits slightly, within minimal targeted modifications due to the XAI guidance. 

\autoref{fig:validity} (b) (bottom) reports, at varying iteration steps, the validity and label preservation rates by the human assessors. The plot shows the average value and the variance between assessors. 
The results from the human assessment differ from those of the automated detector. \tool still outperforms \deepjanus, with its validity and label-preserving rate converging around $0.9$ and $0.7,$ respectively, after $300$ iterations. In comparison, \deepjanus's validity and label preservation rate are lower than that of \tool and exhibit a downward trend. 

\begin{tcolorbox}[boxrule=0pt,sharp corners,boxsep=2pt,left=2pt,right=2pt,top=2.5pt,bottom=2pt]
\begin{center}
\begin{minipage}[t]{0.99\linewidth}
\textbf{RQ\textsubscript{4}}: \textit{
The failure-inducing inputs by \tool are regarded as valid in-distribution inputs, according to state-of-the-art automated input validators. Moreover, they exhibit a high validity rate ($\approx$90\%) and label preservation rate ($\approx$70\%), according to human assessors. 
}
\end{minipage}
\end{center}
\end{tcolorbox}

\subsubsection{RQ\textsubscript{5} (Comparison)}

\autoref{tab:density_coverage} shows, for the various techniques, the results for effectiveness, efficiency, density, and coverage.
In this analysis, we studied the best performing configuration of \tool (\xaiGC in combination with Clustering and Attractor).
\tool outperforms all other gradient-guided methods also in terms of efficiency and effectiveness. DeepXplore generates a low amount of failures since it only inserts noise in the upper-left occlusion window of the image.

\begin{figure}[t]
  \centering

  \includegraphics[width=0.9\linewidth]{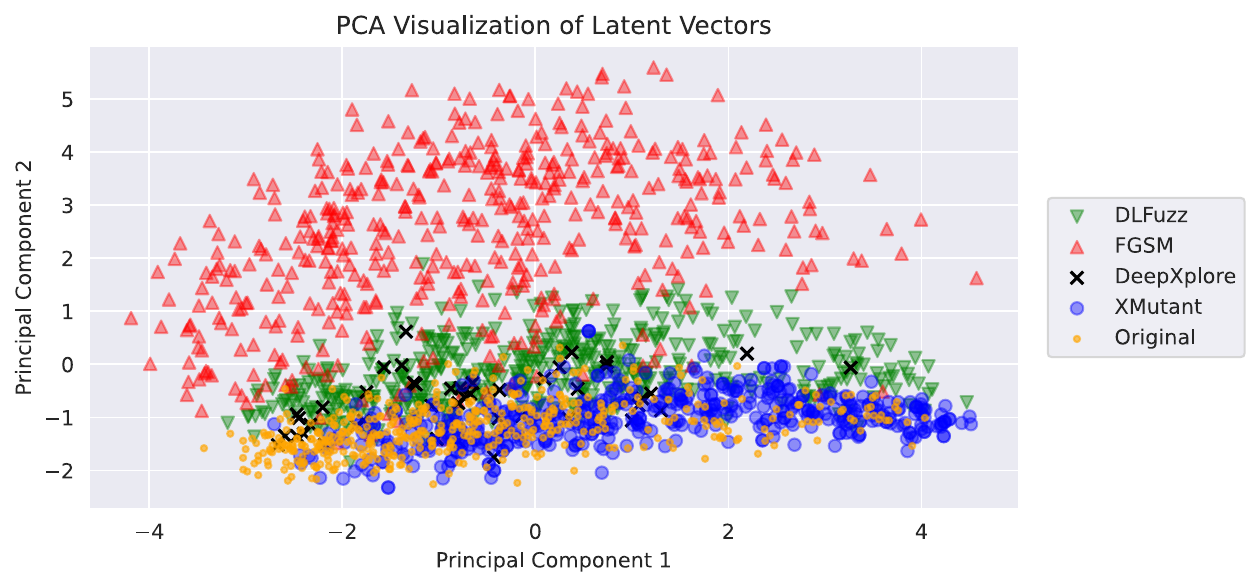}
  \caption{Semantic space visualization of generated inputs by different techniques.}
   \label{fig:pca}
\end{figure}

\autoref{fig:pca} shows that adversarial techniques exhibit a significant shift in the second principal component (PC2) in the PCA space, while the first principal component (PC1) remains relatively aligned with the original data. The failure-inducing inputs generated by \tool generally adhere to the original distribution and fill in sparsely populated regions of the original distribution (towards the right part). PC1 captures global features such as overall brightness and coarse-grained structure, whereas PC2 is more sensitive to local features such as texture details. Despite minor pixel modifications, RIM techniques seem to generate radically different images in the PCA space.
It can be observed that the density and coverage are generally consistent with the PCA visualization, indicating that the inputs generated by the \tool are closest to the original distribution.

\begin{table}[t]
    \centering
    \caption{Comparison of XMutant and gradient-guided adversarial testing
methods.}
    \label{tab:density_coverage}
    \begin{tabular}{lcccc}
        \toprule
        \textbf{Technique} & \textbf{Effectiveness} & \textbf{Efficiency (s)} & \textbf{Density} & \textbf{Coverage} \\
        \midrule
        DLFuzz     & 60.10\% & 0.65  & 0.2156  & 0.34  \\
        FGSM       & 43.80\% & 0.37  & 0.0104  & 0.05   \\
        DeepXplore & 0.02\%  & 76.84 & 0.2591  & 0.11   \\
        \tool      & \textbf{74.15\%} & \textbf{0.33}  & \textbf{0.3824}  & \textbf{0.41}  \\
        \bottomrule
    \end{tabular}
\end{table}

\begin{figure}[h]
  \centering

 \includegraphics[width=\linewidth]{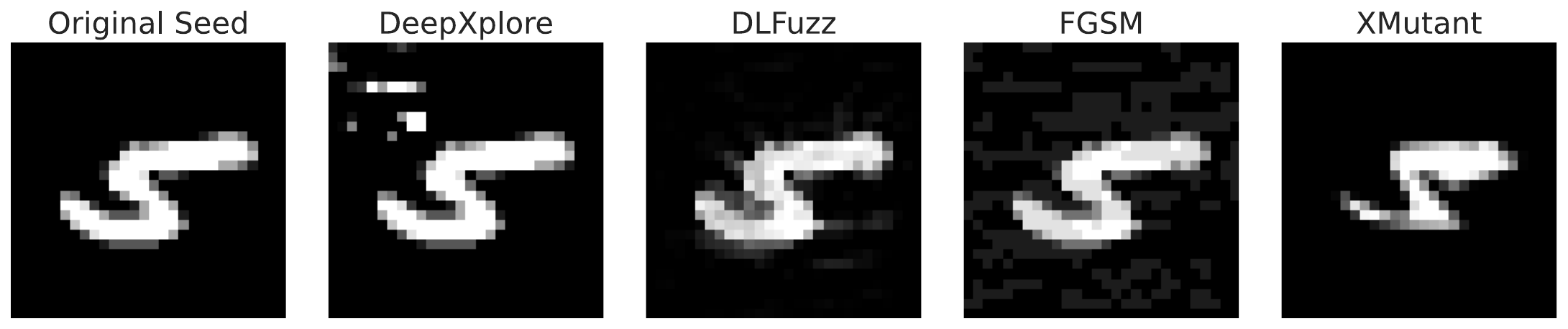}
 \caption{An example of generated failure-inducing inputs by raw input manipulation techniques and \tool.}
   \label{fig:generated_examples}
\end{figure}

\autoref{fig:generated_examples} compares the input produced by \tool and the other three raw input manipulation methods (DLFuzz, FGSM, and DeepXplore). These pixel-level fuzzing approaches inject different patterns of noise, either in the global background or locally focused on some specific areas, while largely preserving the semantic content of the input (the shape of the digit) from the original seed. In contrast, \tool operates on a semantic input representation, modifying the shape of the digit through control-point mutations. This leads to functionally new and visibly distinct test inputs challenging the model under test.

\begin{tcolorbox}[boxrule=0pt,sharp corners,boxsep=2pt,left=2pt,right=2pt,top=2.5pt,bottom=2pt]
\begin{center}
\begin{minipage}[t]{0.99\linewidth}
\textbf{RQ\textsubscript{5}}: \textit{
\tool outperformed 3 gradient-guided adversarial testing methods on four metrics. The failure-inducing inputs by \tool are able to better cover the original distribution in semantic space.
}
\end{minipage}
\end{center}
\end{tcolorbox}

\subsection{Threats to Validity}\label{sec:ttv}

\subsubsection{Internal validity}
We conducted comparisons between all variants of \tool and the baseline within the same experimental framework, using identical benchmarks. A potential internal validity is our implementation of the scripts used to assess these results, which have undergone extensive testing. Additionally, regarding the ADAS model and the simulation platform, we utilized artifacts available from the replication packages of prior studies~\cite{2024-Biagiola-EMSE}. 

\subsubsection{External validity}
The limited number of DL systems in our evaluation poses a threat in terms of the generalizability of our results. To mitigate this, we considered DL systems addressing different tasks, input data, and testing levels. For the local explanation methods, we considered a limited number of XAI methods. 
To address this threat, we selected XAI algorithms from diverse families (e.g., surrogate, gradient, and saliency-based methods). Our results show that \tool consistently outperforms the baseline regardless of the XAI method used. However, it is important to note that the optimal choice of the XAI algorithm is domain-specific. 

\section{Qualitative Analysis}

We analyzed some of the failures qualitatively to understand the reasons why the local explanations are able to  guide the mutation process effectively. We focused our analysis on the digit recognition and advanced driving assistance systems.

\begin{figure}[t]
    \centering
    \begin{subfigure}[b]{0.95\textwidth}
        \centering
        \includegraphics[width=\linewidth]{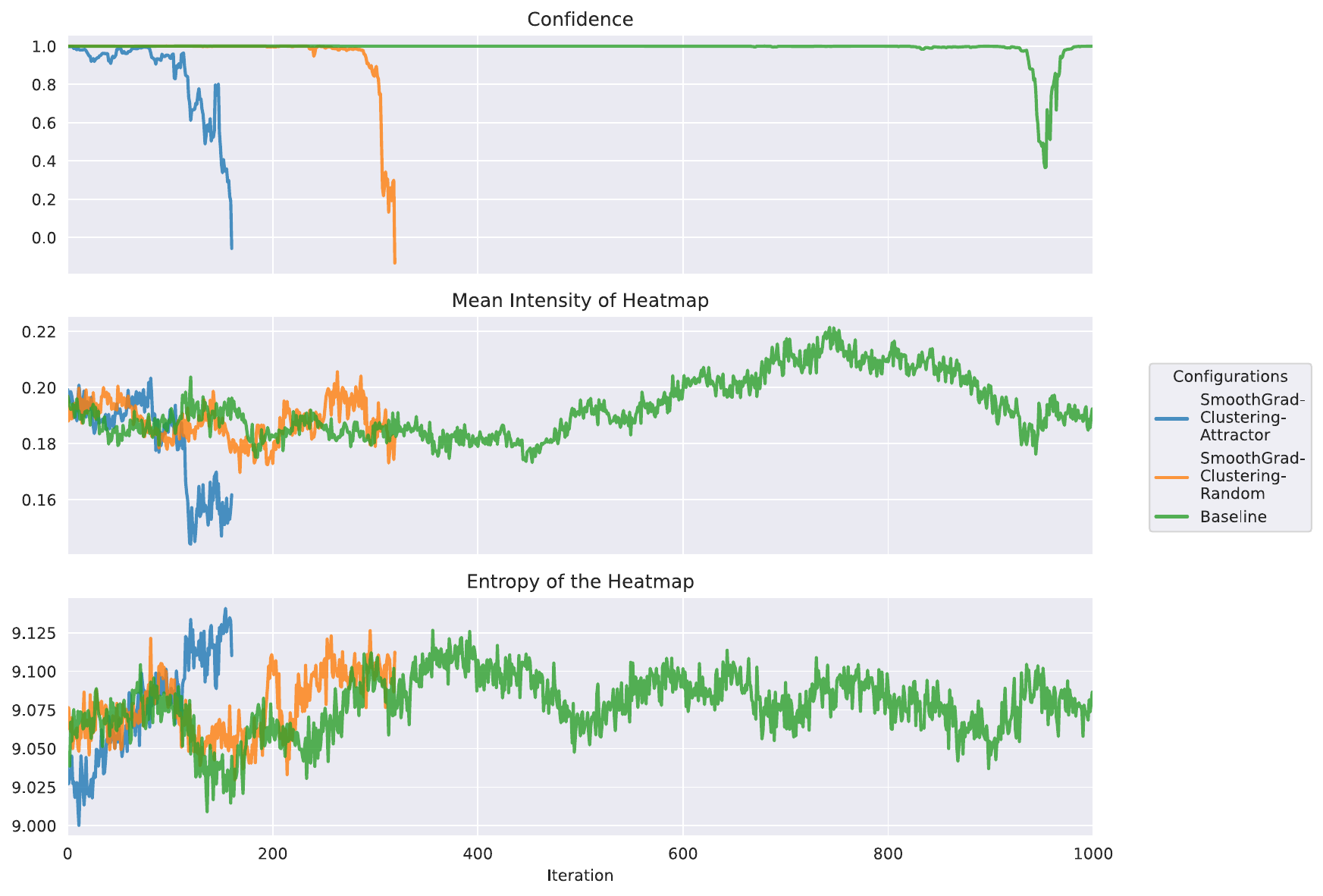}
    \end{subfigure}
    

    \begin{subfigure}[b]{0.95\textwidth}
        \centering
        \includegraphics[width=\linewidth]{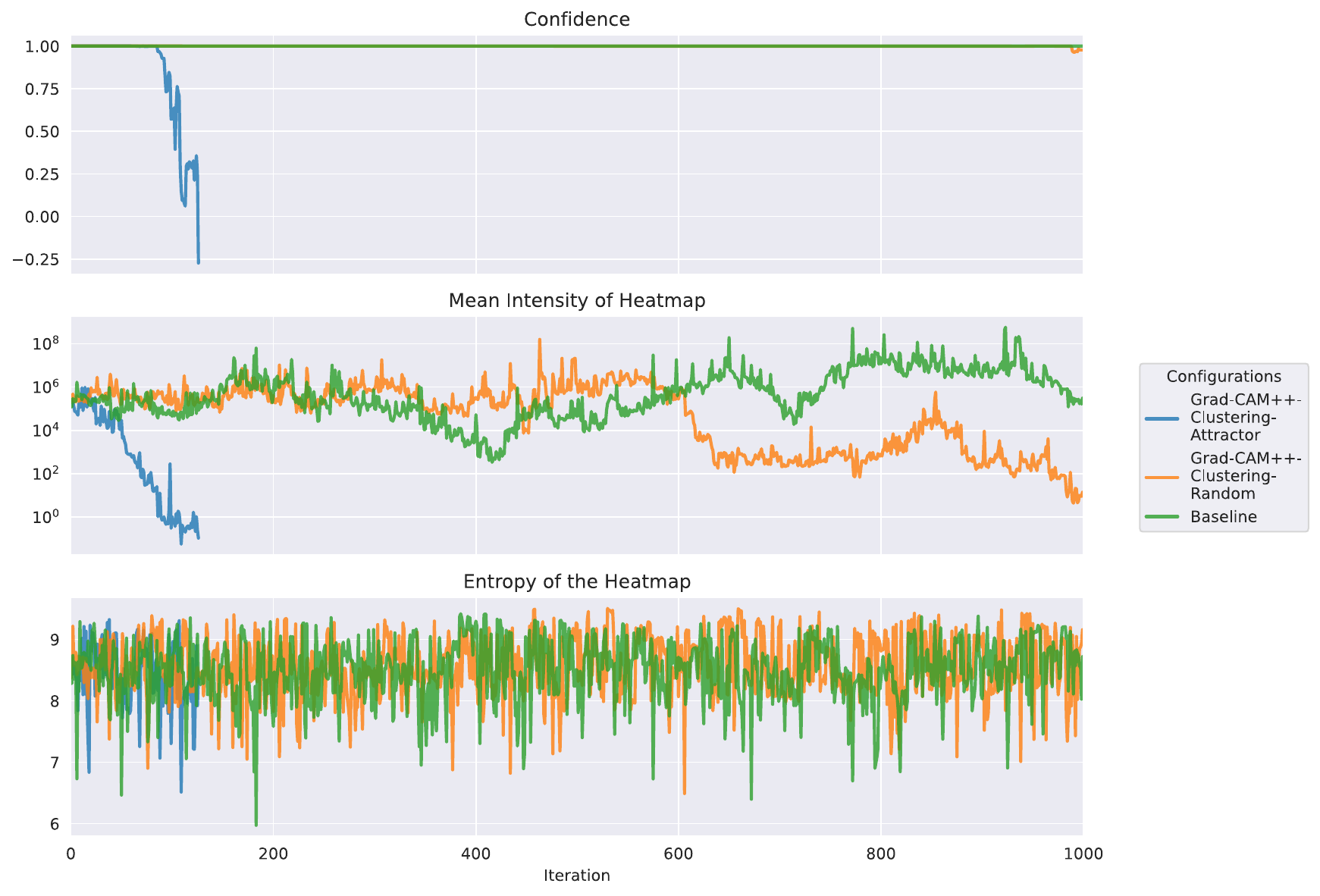}
    \end{subfigure}
    
    \caption{\changed{Qualitative analysis for MNIST by \xaiSG (top) and \xaiGC (bottom).}}
    \label{fig:qualitative}
\end{figure}

\subsection{Qualitative Analysis of Digit Recognition}
For model-level digit recognition, we investigate the underlying relationship between model confidence and heatmaps by observing how they change with different mutation configurations. Since heatmaps are high-dimensional matrices, we employ two metrics to extract relevant information. The first metric is the average intensity, for which we have to remove the normalization applied during local explanation computation to restore the original data. The second metric is Shannon's entropy \cite{shannon1948mathematical}, which captures the level of disorder in the heatmap. A lower entropy value suggests more concentrated or structured intensity patterns, whereas a higher entropy value indicates a more uniform distribution of intensity values. The first metric reflects the raw intensity comparison across a series of heatmaps, while the second metric quantifies internal variations within a single heatmap.

We analyzed two XAI methods, \xaiSG and \xaiGC (\autoref{fig:qualitative}), selecting three mutation configurations guided by the same seed. The first observation is related to \textbf{gradient attenuation}, as the mean intensity of both methods decreases with the confidence, especially those driven by \tool. Notably, \xaiGC's mean intensity drops significantly, thus it is visualized in a logarithmic scale. Since \xaiSG reflects gradients across the entire model, the gradient attenuation is less pronounced w.r.t. \xaiGC, which instead focuses on local gradients at the last convolutional layer, directly exposing the gradient attenuation in this layer and resulting in a more pronounced decline.
Our second observation is related to \textbf{entropy degradation}. The inputs generated by \tool cause the entropy to increase when the confidence decreases, indicating that the DNN progressively loses its focus. This phenomenon is also more observable with the mutation guided by \tool equipped with \xaiSG, since it generates noise-averaged gradients, whereas the entropy of heatmap generated by \xaiGC appears noisy.

\subsection{Qualitative Analysis of ADAS}
For system-level advanced driving assistance, we considered three configurations of \tool---\xaiGC (Random), \xaiGC (High)---and further analyzed the driving quality degradation before/after mutation.

Additionally, to allow a fine-grained comparison across approaches for ADAS, we assess the driving quality degradation (DQD) of the lane-keeping model using two metrics~\cite{2021-Jahangirova-ICST}: (1)~the maximum CTE value (cross-track error, i.e., the lateral position of the vehicle w.r.t. the center road) and (2)~the Euclidean norm of the steering angle output in the segment of the road affected by the mutation. 
We record the driving data on the road segment relevant to the candidate control point and derive the driving quality value by applying the selected norm to the raw driving data. Consequently, we calculate DQD by subtracting the pre-mutation driving quality from the post-mutation driving quality.

For each technique, we considered 6,622 pairs of pre- and post-mutation driving quality quantities, involving the maximum CTE value and the Euclidean distance of steering angle in the mutated road segments.

\begin{figure}[t]
  \centering
  \includegraphics[width=\linewidth]{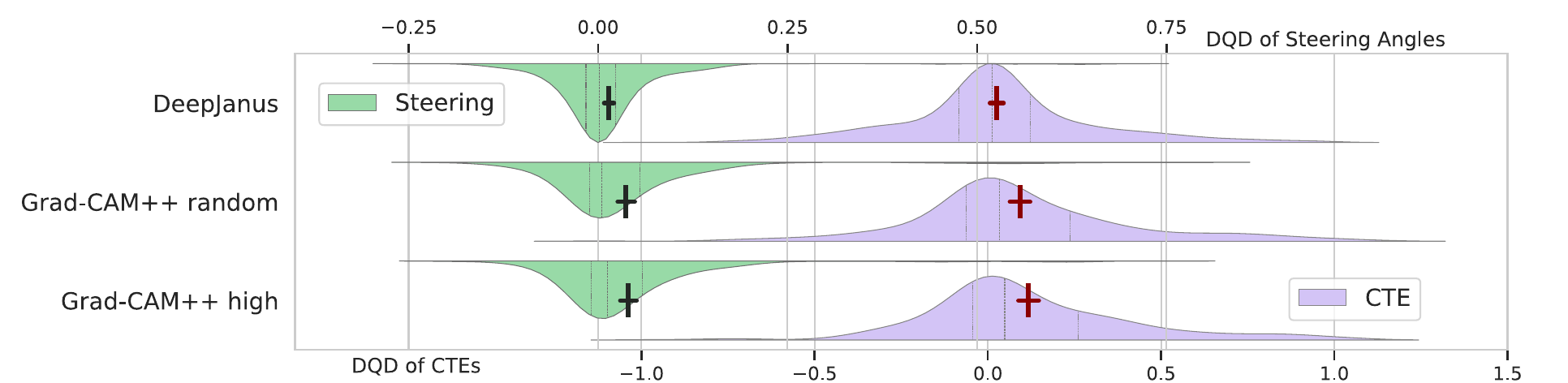}
  \caption{Distributions of Driving Quality Degradation (DQD) based on CTEs/steering angles.}
   \label{fig:dqd}
\end{figure}

\autoref{fig:dqd} depicts the distribution of DQD, i.e., the subtraction of driving quality pairs. \deepjanus produces only minor variations, with smaller means and a tighter distribution, with its median closely aligned with symmetry around zero. On the other hand, the mutations led by \tool result in larger mean and variance values for DQD, as well as more data points distributed in areas with high values, indicating a higher driving quality degradation due to more challenging road topologies. Within each approach, we conduct the above-mentioned statistical test over the driving quality distributions before and after mutation. 
The differences in the quality of driving before/after mutation were found to be statistically significant according to the non-parametric Mann-Whitney U test~\cite{Wilcoxon1945} ($p$-value $<$ 0.01), which is not the case for the \deepjanus approach.

\section{Discussion}\label{sec:discussion}

\head{XAI for DL Test Generation}  
In common practice, XAI's explanations are used qualitatively by humans to understand how a DNN processes the inputs. However, in our research, we use them quantitatively, under the assumption that they contain relevant information related to the behavior of DNNs~\cite{samek2017explainable,abs-2201-00009}.
Our experiments showed the effectiveness brought by XAI in guiding semantic-based fuzzing, as focusing on the area of attention led to faster discovery of misbehaviors while retaining their validity, as the mutations are more focused and less disruptive. 

Although different XAI methods may vary significantly in their quality and behavior, particularly in terms of established evaluation metrics such as faithfulness and sensitivity~\cite{li2021experimental, adebayo2018sanity}, our results demonstrate that \tool consistently improves failure discovery across most configurations. This suggests that \tool does not require the explanations to be fully accurate or faithful, but rather benefits from the heuristic signal provided by these explanations to increase the likelihood of failure-inducing inputs. 
This is due to the nature of fuzzing itself, which is inherently exploratory and tolerant of imperfect guidance. Therefore, \tool remains robust even when the underlying XAI techniques vary in their quality and behavior.

Additionally, understanding the characteristics of these explanations by different XAI methods is crucial for \tool to leverage them for effective mutation selection and direction. 
For example, attribution-based methods are particularly useful for determining mutation direction, while sensitivity-based methods are helpful to guide selection toward more influential input components. \tool adapts its mutation strategy accordingly to accommodate these differences in explanation semantics.

The results of this study showed the significant benefits of utilizing XAI for semantic-based fuzzing. XAI enables the identification of DNN's misbehaviors while ensuring that the tests remain valid and relevant to the original input domain, thanks to the semantic representation and focused test generation. 
Although the generation of local explanation introduces additional computational overhead, our findings suggest that the trade-off is justified. Our effectiveness results, even for complex applications such as advanced driving assistance systems, indicate that the benefits of XAI in failure exposure outweigh the costs of computing heatmaps. 
Our experiments suggest that incorporating XAI (except for the model-agnostic XAI method \xaiLime) into the testing workflow is beneficial. Additionally, the adopted open-source XAI methods can be further optimized for efficiency by parallelizing multiple backpropagations or DNN inferences. We will evaluate this aspect in our future work.

\head{\tool's Configurations}
All configurations of \tool are stable in terms of effectiveness across all iterations but our study shows that the choice of the optimal XAI algorithm is task-specific. 

For model-level sentiment analysis, both \xaiLime and \xaiIG can generate a significant amount of failures, as their explanations differentiate between the positive and negative impacts of semantic concepts, allowing for applying targeted mutations. However, due to \xaiLime's high computational cost, \xaiIG is the most suitable candidate for \tool in practice for textual inputs.
For model-level digit classification, all XAI algorithms are effective, with SmoothGrad (Clustering Attractor) showing the best results. We believe this is associated with its more direct method of reflecting gradient information within the input space, which significantly influences the DNN's outcome during model-level testing.
For system-level automated driving assistance, our findings indicate that \xaiIG did not provide benefits compared to the baseline. Conversely, CAM-based algorithms proved highly effective and efficient if a low testing budget was available. However, in system-level testing, it is challenging to identify a reasonable direction for mutation, due to cumulative uncertainty and flakiness of the system~\cite{10.1007/s10664-023-10433-5}. This results in targeted mutations (High versus Random) being less effective and offering limited guidance since 
the system may exhibit different behaviors across consecutive test runs.

\head{XAI-guided Tests preserve validity and original distribution semantics}
Thanks to the usage of semantic-based input representation (i.e., a model), \tool generates failure-inducing inputs while retaining the essential properties of validity and label accuracy. As a result, the generated inputs show a more natural distribution in the embedding space while exhibiting \textit{functional novelty}, serving as an effective mean to evaluate the model's generalization ability.
In summary, our initial exploration into the usage of XAI for enhancing semantic-based test generation has proved very promising. By carefully selecting and applying the most appropriate XAI methods, it is possible to significantly improve the efficiency and effectiveness of the testing process of complex DL systems.

\section{Related Work}\label{sec:related}

With the increasing application of DL to safety-critical domains such as autonomous during, XAI algorithms represent one of the default options to debug the predictions and failures of a DL system. In this section, we focus on the main related propositions.

\subsection{Test Generation for Deep Learning Systems}

The three main families of DL test generation are semantic-based input representation, raw input manipulation, and latent space manipulation. 

Semantic Input Manipulation (SIM) techniques leverage a semantic representation of the input domain (e.g., a model) to generate test inputs, similar to conventional model-driven engineering practices that uphold compliance with domain-specific constraints~\cite{Abdessalem-ASE18-1,Abdessalem-ASE18-2,Abdessalem-ICSE18,Gambi:2019:ATS:3293882.3330566,2020-Riccio-FSE,isa,2023-Stocco-EMSE,2023-Stocco-TSE}. 
The manipulation occurs on the model, which is subsequently reconverted to the original format~\cite{larman1998applying}.
SIM techniques operate within a restricted input space, specifically the control parameters of the model representation. These techniques enhance the realism of the produced outputs by implementing appropriate model constraints.

Several search-based SIM approaches have been applied to DL-based image classifiers. DeepHyperion~\cite{zohdinasab2021deephyperion} uses the MAP-Elites Illumination Search algorithm~\cite{DBLP:journals/corr/MouretC15} to explore the feature space of the input domain and identify misbehavior-inducing features. DeepMetis~\cite{2021-Riccio-ASE} a SIM approach that generates inputs that behave correctly on original DL models and misbehave on mutants obtained through injection of realistic faults~\cite{2020-Humbatova-ICSE}, which can be useful to enhance the mutation killing ability of a test set.
\deepjanus~\cite{2020-Riccio-FSE} is the SIM approach most related to this work since it performs model-based testing of DL systems. Therefore, we performed an explicit empirical comparison with the \deepjanus approach in this work.

To the best of our knowledge, no XAI information is used in existing semantic-based tools to guide test generation. As such, they can be used in conjunction with \tool to improve their effectiveness.

Raw input manipulation (RIM) techniques involve modifying an image's original pixel space to create a new input by perturbing the pixel values~\cite{lambertenghi_ICST25}. 
RIM techniques aim to produce minimal, often imperceptible changes to original to trigger misbehaviors in the DL system. These methods target different aspects of testing, such as data augmentation or adversarial attacks, which are not directly aligned with our goal. Our method is a \textit{functional} test generator, differing from adversarial testing in both goals and techniques. Functional testing creates new, valid, in-distribution inputs to evaluate a DNN's generalization. In contrast, adversarial testing adds minor perturbations to original inputs to test \textit{robustness}. 
However, for completeness, we describe the main propositions next.

DeepXplore~\cite{pei2017deepxplore} employs various techniques, including occlusion, light manipulation, and blackout to cause misbehaviors. These perturbations are intended to improve neuron coverage within the DL system.
DLFuzz~\cite{guo2018dlfuzz} introduces noise to the seed image to increase the likelihood of system misbehavior. DLFuzz generates adversarial inputs for DL systems without relying on cross-referencing other similar DL systems or manual labeling. 
DeepTest~\cite{deeptest} alters the images using synthetic affine transformation from the computer vision domain, such as blurring and brightness adjustments, to create simulated rain/fog effects.

Differently, our technique targets functional testing, specifically testing for generalization of DL systems. We achieve this by using XAI guidance in the pixel space, and link it to the model input to generate inputs beyond the original datasets, while remaining within the same distribution. {In \autoref{sec:study}, we have showed that our technique generates more natural samples that adhere to the original data's manifold, whereas these RIM methods, despite their minimal pixel perturbations, artificially inflate failure rates via distribution shifts.

Latent space manipulation (LIM) techniques generate new inputs by learning and reconstructing the underlying distribution of the input data~\cite{2025-Guo-arxiv}. 
Sinvad~\cite{kang2020sinvad} constructs the input space using Variational Autoencoders (VAE)~\cite{kingma2013auto, doersch2016tutorial} and navigates the latent space by adding a random value sampled from a normal distribution to a single element of the latent vector. Sinvad aims to explore the latent space by maximizing either the probability of misbehaviors, estimated from the softmax layer output, or by surprise coverage~\cite{kim2019guiding}. 
The Feature Perturbations technique~\cite{dunn2021exposing, DBLP:journals/corr/abs-2001-11055} involves injecting perturbations into the output of the generative model’s first layers, which represent high-level features of images. These perturbations can affect various characteristics of the image, such as shape, location, texture, or color. 
DeepRoad~\cite{deeproad} generates driving images using Generative Adversarial Networks (GANs)~\cite{goodfellow2014generative, goodfellow2020generative} for image-to-image translation.
CIT4DNN~\cite{dola2024cit4dnn} combines VAE and combinatorial testing~\cite{cit605761}. This allows the systematic exploration and generation of diverse and infrequent input datasets. 
CIT4DNN partitions latent spaces to create test sets that contain a wide range of feature combinations and rare occurrences. Instance Space Analysis, aims to pinpoint the critical features of test scenarios that impact the detection of unsafe behavior~\cite{isa}. Mimicry~\cite{weissl2025targeted} explores the behavioral boundaries of deep learning classifiers by generating boundary inputs through manipulation of disentangled latent representations learned by a style-based generative model.

These aforementioned LIM techniques require large-scale training data to capture complete feature distributions, which inherently limits their applicability in resource-constrained or open-world settings. For non-stationary data streams (e.g., evolving NLP token distributions or evolving self-driving scenarios), LIM requires frequent retraining to accommodate changes in the distribution, which incurs excessive computational costs. Indeed, existing testing propositions are limited to DL systems that take as input individual images~\cite{isa,dola2024cit4dnn,deeproad,2025-Maryam-ICST,2025-Baresi-ICSE}.
Differently from these LIM approaches, \tool leverages the local explanation from existing seeds to guide model-based test generation, by perturbing semantic representations instead of latent vectors from a learned manifold. In our paper, we showed that the semantic-based representations made \tool applicable across different case studies, both at the model and at the system level. 

\subsection{XAI for DL Testing}

Zohdinasab et al.~\cite{10304866} compare three state-of-the-art techniques for explanation of DL failures. They show that local and global XAI techniques provide dissimilar explanations for the same inputs and further research is needed to produce better explanations.
VisualBackProp~\cite{05418} was created to visualize which group of pixels of the input image contributes more to the predictions of a convolutional neural network. Kim and Canny~\cite{10631} explore the use of heatmaps for explaining the CNN behavior in an ADS.
Xu et al.~\cite{09405} investigated the use of XAI techniques to detect action-inducing objects, i.e., objects that have a relevant effect on a driving decision, and jointly predict actions and their respective explanations. 
ThirdEye~\cite{2022-Stocco-ASE} focuses on failure prediction of lane-keeping autonomous driving systems during hazardous driving conditions. They turn heatmaps into confidence scores that are used to discriminate safe from unsafe driving behaviors. The intuition is that uncommon heatmaps are associated with unexpected runtime conditions. In this work we use ThirdEye, equipped with the heatmap derivative configuration, to provide such scores during system-level testing. 

Fahmy et al.~\cite{DBLP:journals/corr/abs-2002-00863} apply clustering to heatmaps capturing the relevance of the DNN predictions to automatically support the identification of failure-inducing inputs. Such data is used for the retraining of a gaze detection system that uses DNNs to determine the gaze direction of the driver. The authors present an extension of the previous work~\cite{arxiv.2204.00480} in which inputs identified by the heatmap-based mechanism are given in input to a search-based test generator. 

In contrast, in this work, we use local explanations to support the early detection of corner case inputs of a DL model. In our work we experiment with different XAI algorithms, showing that the choice of the best algorithm is domain and testing level dependent.
\section{Conclusions}\label{sec:conclusions}

In this work, we describe and evaluate \tool, a semantic-based fuzzer for DL systems that generates inputs that focus on the attention of the system under test through mutations that are informed by the local explanation available from XAI algorithms. We evaluated \tool on both model-level and system-level testing. Our empirical studies show that \tool is significantly more effective and efficient  
than the state-of-the-art test generation approaches,
 while preserving a high validity rate of failing test inputs.

\section{Data Availability}\label{sec:da}

All our results, the source code, and the simulator are accessible and can be reproduced~\cite{replication-package}. 

\newpage
\section{Appendix}\label{sec:appendix}

\subsection{Effect of Window Size}

To validate the choice of the window size $ws=3$ for the \tool equipped with Square-Window in MNIST study, we conducted an ablation study across different values (1, 3, 5, 7), as shown in \autoref{fig:ws}. We report the failure rates using both \xaiSG and \xaiGC, for 2000 seeds and with the budget of 200 iterations.

\begin{figure}[h]
    \centering
    \includegraphics[width=1\linewidth]{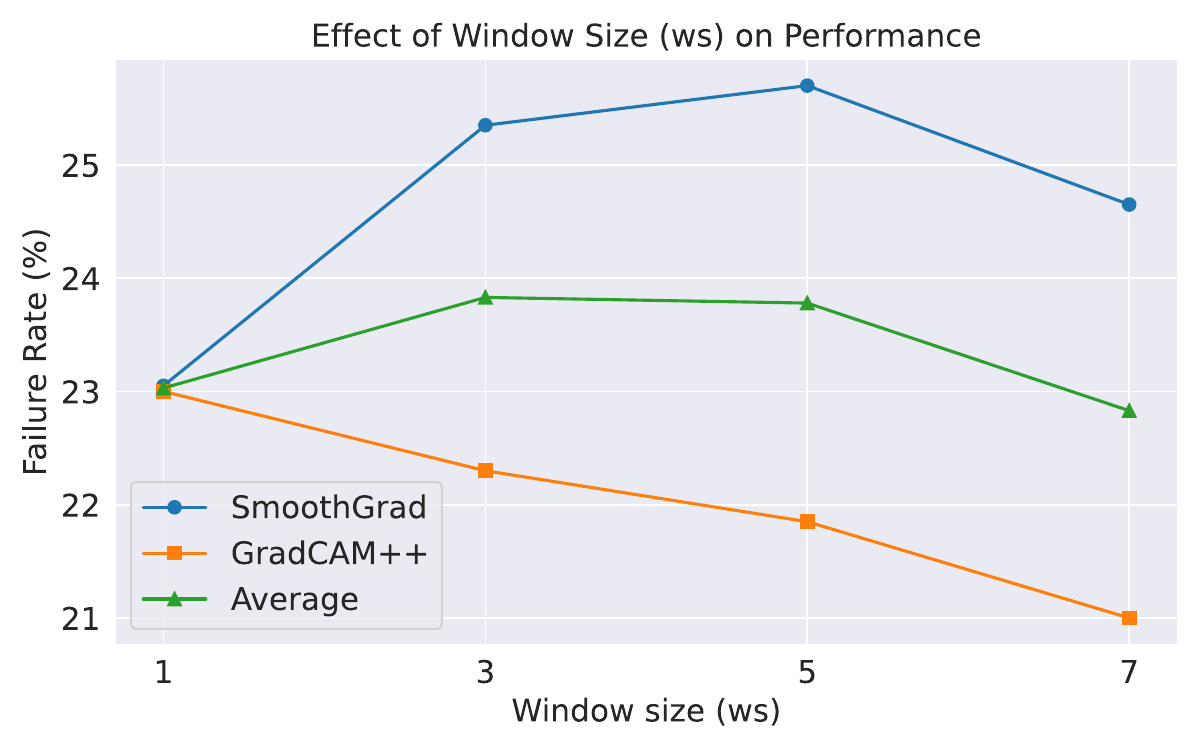} 
    \caption{Effect on Window Size }
    \label{fig:ws}
\end{figure}

While \xaiSG peaks around $ws=5$, \xaiGC performs best at $ws=1$. On average, $ws=3$ yields the highest combined score across methods. However, the variation is relatively minor (within ~2\%), suggesting the method is not overly sensitive to this parameter. Based on this, we selected $ws=3$ for running our experiments to balance performance and generality.

\subsection{Comprehensive result for RQ\textsubscript{1}, RQ\textsubscript{2}, and RQ\textsubscript{3}}

\begin{table*}[h]
\renewcommand{\arraystretch}{1}
\setlength{\tabcolsep}{3.5pt}

\centering

\scriptsize

\caption{Detailed results for RQ\textsubscript{1}, RQ\textsubscript{2} and RQ\textsubscript{3}. For each configuration, the two values indicate the \textit{Cumulative Failure Rate [\%]/ Relative Efficiency} over different Iterations. P-values indicating statistically significant differences are highlighted in bold.}

\resizebox{\columnwidth}{!}{%

\begin{tabular}{c@{\hspace{20pt}}c@{\hspace{36pt}}c@{\hspace{46pt}}c@{\hspace{45pt}}c@{\hspace{45pt}}c}

\toprule

&Iterations & \xaiSG & \xaiLime & \xaiIG & DJ \\ 
\midrule
& 10	& 14/1.93	& 11/1.44 & 16/2.05		& 7/- \\
& 20	& 22/2.08	& 32/2.31 & 31/2.71		& 10/- \\
& 30	&  31/2.08	& 42/2.74 & 46/2.92		& 14/- \\
& 40	& 39/2.12	& 51/2.82 & 57/3.05		& 18/- \\
& 50	& 41/2.12	& 60/2.86 & 64/3.09		& 21/- \\
& 60	& 47/2.08	& 63/2.83 & 67/3.04		& 24/- \\
& 70	& 49/2.03	& 66/2.75 & 70/2.94		& 27/- \\
& 80	& 51/1.98	& 70/2.69 & 74/2.87  	& 29/- \\
& 90	& 53/1.93	& 72/2.64 & 76/2.80		& 31/- \\
\multirow{-10}{*}{\rotatebox[origin=c]{90}{IMDB}} & 100   & 56/1.90	& 76/2.74	& 76/2.59	& 32/-\\

\midrule
						
& { p-value } &\bf 6.93E-18  & \bf 1.17E-17 &\bf 7.90E-18 & -\\
\multirow{-2}{*}{{RQ\textsubscript{1}}} & {effect size} & large & large & large & -\\

\midrule

& { p-value } &\bf 2.53e-04  &\bf 1.20e-06 &\bf 2.92e-06 & -\\
\multirow{-2}{*}{{RQ\textsubscript{2}}} &{effect size} & medium & large & large & -\\

\end{tabular}
}

\resizebox{\textwidth}{!}{
\begin{tabular}{cccccccccccc}

\toprule

\multicolumn{2}{c}{} & \multicolumn{3}{c}{\xaiSG} & \multicolumn{3}{c}{\xaiGC} 
& \multicolumn{3}{c}{\xaiIG} & {DJ} \\
\cmidrule(r){3-5}
\cmidrule(r){6-8}
\cmidrule(r){9-11}

\multicolumn{2}{c}{} & \multicolumn{2}{c}{Clustering} & \multicolumn{1}{c}{SW} & \multicolumn{2}{c}{Clustering} & \multicolumn{1}{c}{SW} 
& \multicolumn{2}{c}{Clustering} & \multicolumn{1}{c}{SW} & - \\

\cmidrule(r){3-4}
\cmidrule(r){5-5}
\cmidrule(r){6-7}
\cmidrule(r){8-8}
\cmidrule(r){9-10}
\cmidrule(r){11-11}

\multicolumn{2}{c}{{Iterations}} & Attractor & {Random} & {Random} & Attractor & {Random} & {Random} 
& Attractor & {Random} & {Random} & - \\ 

\midrule


 & {100} & { 44/7.00} & 11/1.83 & {11/2.11} & 36/4.81 & 07/1.03 & {09/1.32} & 34/5.54 & 13/2.06 & {13/2.15} & 06/- \\
 & {200} & { 77/6.37} & 24/1.77 & {24/1.82} & 74/5.54 & 21/1.32 & {22/1.48} & 60/4.92 & 27/2.00 & {26/1.97} & 15/- \\
 & {300} & { 88/4.97} & 38/1.62 & {36/1.60} & 88/4.61 & 35/1.36 & {34/1.41} & 74/3.92 & 41/1.80 & {37/1.72} & 26/- \\
 & {400} & 94/4.03 & 47/1.52 & {45/1.47} & 94/3.84 & 46/1.34 & {45/1.36} & 80/3.26 & 51/1.67 & {47/1.57} & 35/- \\
 & {500} & {96/3.42} & 55/1.44 & {53/1.39} & {97/3.30} & 54/1.32 & {54/1.32} & 85/2.82 & 59/1.57 & {54/1.47} & 43/- \\
 & {600} & {97/2.99} & 62/1.38 & {60/1.33} & {98/2.91} & 60/1.28 & {61/1.29} & 87/2.51 & 66/1.49 & {61/1.39} & 51/- \\
 & {700} & {98/2.67} & 67/1.32 & {65/1.28} & {98/2.61} & 66/1.25 & {67/1.26}& 89/2.27 & 71/1.43 & {66/1.33} & 58/- \\
 & {800} & {98/2.43} & 71/1.28 & {69/1.25} & {99/2.39} & 70/1.22 & {73/1.24} & 90/2.10 & 76/1.38 & {71/1.29} & 62/- \\
 & {900} & {99/2.26} & 75/1.25 & {73/1.22} & {99/2.23} & 74/1.20 & {77/1.22} & 91/1.96 & 79/1.35 & {74/1.26} & 67/- \\
\multirow{-10}{*}{\rotatebox[origin=c]{90}{MNIST}} & {1000} & {99/2.12} & 78/1.23 & {76/1.20} & {99/2.09} & 77/1.19 & {80/1.21} & 92/1.86 & 82/1.32 & {77/1.23} & 70/- \\ 

\midrule
						
& {p-value } & {\bf  4.94E-165} & {\bf 1.09E-164} & {\bf 3.46E-165} & {\bf 2.57E-164} & {\bf 1.19E-161} &{ \bf 2.10E-164} & {\bf 4.86E-165} & {\bf 7.03E-165} &{\bf  7.05E-165} & - \\ 

\multirow{-2}{*}{{RQ\textsubscript{1}}} & {effect size} & {large} & {small} & {small} & {large} & {small} & {small} & {large} & {medium} &{small} & - \\

\midrule
						
& {p-value } & {\bf  1.75E-183} & {\bf 2.47E-15} & {\bf 1.59E-10} & {\bf 4.46E-188} & {\bf 1.55E-10} &{ \bf2.55E-08} & {\bf 2.06E-142} & {\bf 2.05E-25} &{\bf  3.91E-15} & - \\ 

\multirow{-2}{*}{{RQ\textsubscript{2}}} & {effect size} & {large} & {small} & {small} & {large} & {small} & {negligible} & {large} & {small} &{small} & - \\

\toprule

\multicolumn{2}{c}{} & \multicolumn{3}{c}{\xaiSG} & \multicolumn{3}{c}{\xaiGC}  & \multicolumn{3}{c}{\xaiIG} & {DJ} \\

\cmidrule(r){3-5}
\cmidrule(r){6-8}
\cmidrule(r){9-11}

\multicolumn{2}{c}{Iterations} & {Random} & {Low} & {High} & {Random} & {Low} & {High} & {Random} & {Low} & {High} & {-} \\ \midrule
 & {{ 5}} & { 27/2.00} & { 31/2.20} & {{ 42/2.87}} & { 49/2.80} & { 42/2.53} & {{ 36/2.07}}  & { 45/2.13} & { 36/2.53} & {{ 40/2.60}} & { 22/-} \\
 & {{ 10}} & { 65/1.35} & { 60/1.36} & {{ 67/1.69}} & { 75/1.86} & { 87/1.78} & {{ 75/1.60}}  & { 64/1.57} & { 65/1.60} & {{ 58/1.56}} & { 51/-} \\
 & {{ 15}} & { 84/1.33} & { 76/1.25} & {{ 82/1.45}} & { 85/1.58} & { 93/1.65} & {{ {95/1.54}}}  & { 75/1.38} & { 78/1.37} & {{ 80/1.38}} & { 65/-} \\
 & {{ 20}} & { 89/1.31} & { 82/1.21} & {{ 84/1.36}} & { {95/1.47}} & { 93/1.53} & {{ {95/1.48}}}  & { 87/1.30} & { 85/1.30} & {{ 85/1.32}} & { 75/-} \\
 & {{ 25}} & { 91/1.25} & { 85/1.16} & {{ 93/1.29}} & { {95/1.39}} & { {95/1.42}} & {{ {95/1.39}}}  & { 91/1.25} & { 87/1.23} & {{ 89/1.25}} & { 84/-} \\
& {{ 30}} & { 91/1.21} & { 89/1.13} & {{ {95/1.24}}} & { {95/1.32}} & { {95/1.35}} & {{ {95/1.32}}}  & { {95/1.21}} & { 87/1.18} & {{ 89/1.20}} & { 84/-} \\
& {{ 35}} & { 93/1.18} & { 89/1.11} & {{ {95/1.21}}} & { {95/1.27}} & { {95/1.29}} & {{ {95/1.27}}}  & { {95/1.19}} & { 91/1.15} & {{ 91/1.17}} & { 89/-} \\
\multirow{-8}{*}{\rotatebox[origin=c]{90}{ADAS}} & {{ 40}} & { {95/1.15}} & { 89/1.09} & {{ {95/1.18}}} & { {95/1.23}} & { {95/1.25}} & {{ {95/1.23}}} & { {95/1.16}} & { 93/1.12} & {{ {95/1.14}}} & { {95/-}} \\ 

\midrule

&{p-value} & {\bf 8.23E-08} & {\bf 5.26E-06 } & {\bf 1.05E-07} & {\bf 1.07E-07} & {\bf 1.10E-07} &{\bf  1.09E-07} & {\bf 1.05E-07} & {\bf 1.54E-07} &{\bf 1.07E-07} & -\\
\multirow{-2}{*}{{RQ\textsubscript{1}}} & {effect size} & {small} & {small} & {small} & {medium} & {medium} &{medium} & {small} & {small} &{small} & - \\

\midrule

&{p-value} & {\bf  9.62E-03} & {2.87E-01} & {\bf 6.93E-03} & {\bf 1.77E-04} & {\bf 1.33E-04} &{ \bf 2.19E-04} & {6.05E-02} & {5.33E-02} &{  1.47E-02} & -\\

\multirow{-2}{*}{{RQ\textsubscript{2}}} & {effect size} & {small} & {-} & {small} & {medium} & {medium} &{medium} & {-} & {-} &{-} & - \\

\bottomrule

\end{tabular}
} 
\label{tab:result}
\end{table*}

\autoref{tab:result} comprehensively presents the effectiveness, performance, and comparative results for all configurations of \tool and \deepjanus as the baseline, for all case studies (IMDB, MNIST, and ADAS).
The table reports the information at different iteration intervals, respectively, 10 for IMDB, 100 for MNIST, and 5 for ADAS. 
For each configuration of \tool and \deepjanus, the first number denotes the cumulative failure rate, whereas the second number indicates the relative efficiency with respect to the baseline. 

\subsection{Sanity Check on LLM's Validation}

We conducted a sanity check of ChatGPT's output by randomly sampling 18 generated test cases across different configurations. Each case included the modified input, ChatGPT's sentiment prediction, and its explanation. Here is an example of generated input with the labels provided by SUT and ChatGPT's validation.

\begin{verbatim}
Generated Text:
"Unreal and contrived melodrama, with a screenplay altered and altered 
by Mel from his play, The Man Business Organization. A room put up and
improving <unk>, taking atomic and atomic amount il, a troubled and 
trouble oneself <unk>, who may be homicidal and murderous...",

Expected Label: "Negative",
Predicted Label: "Positive",
ChatGPT's Answers: "negative. The review expresses a strong sense of 
dissatisfaction with the film, using words like 'unreal', 'contrived'... 
These terms indicate a negative sentiment towards the movie, suggesting 
that the reviewer found it lacking in quality and engaging elements."
\end{verbatim}

5 human annotators independently assessed whether (i) the predicted sentiment aligned with the input and (ii) the explanation was logically consistent and specific.
To quantify the overall agreement among human participants and ChatGPT, we computed Fleiss' kappa, obtaining a value of 0.71, 
indicating a substantial inter-rater agreement~\cite{Fleiss:1971,Landis:1977}.
In addition to the group-level measure, we assessed the individual consistency of each participant by comparing their responses with those of the others on every task. The results reveal that the assessment of ChatGPT was the most reliable (0.88) with the highest agreement rate, closely followed by other human annotators (0.78-0.88). 

\newpage 

\balance
\bibliographystyle{spmpsci}
\bibliography{paper}

@inproceedings{zhang2024enhancing,
	title        = {Enhancing Valid Test Input Generation with Distribution Awareness for Deep Neural Networks},
	author       = {Zhang, Jingyu and Keung, Jacky and Ma, Xiaoxue and Li, Xiangyu and Xiao, Yan and Li, Yishu and Chan, Wing Kwong},
	year         = 2024,
	booktitle    = {2024 IEEE 48th Annual Computers, Software, and Applications Conference (COMPSAC)},
	pages        = {1095--1100},
	organization = {IEEE}
}

@misc{2025-Guo-arxiv,
      title={Foundation Models in Autonomous Driving: A Survey on Scenario Generation and Scenario Analysis}, 
      author={Yuan Gao and Mattia Piccinini and Yuchen Zhang and Dingrui Wang and Korbinian Moller and Roberto Brusnicki and Baha Zarrouki and Alessio Gambi and Jan Frederik Totz and Kai Storms and Steven Peters and Andrea Stocco and Bassam Alrifaee and Marco Pavone and Johannes Betz},
      year={2025},
      eprint={2506.11526},
      archivePrefix={arXiv},
      primaryClass={cs.RO},
      url={https://arxiv.org/abs/2506.11526},
}

@article{weissl2025targeted,
	title        = {Targeted Deep Learning System Boundary Testing},
	author       = {Wei\ss{}l, Oliver and Abdellatif, Amr and Chen, Xingcheng and Merabishvili, Giorgi and Riccio, Vincenzo and Kacianka, Severin and Stocco, Andrea},
	year         = 2025,
	month        = oct,
	journal      = {ACM Trans. Softw. Eng. Methodol.},
	publisher    = {Association for Computing Machinery},
	address      = {New York, NY, USA},
	doi          = {10.1145/3771557},
	issn         = {1049-331X},
	url          = {https://doi.org/10.1145/3771557}
}

@article{jiang2024validity,
	title        = {Validity Matters: Uncertainty-Guided Testing of Deep Neural Networks},
	author       = {Jiang, Zhouxian and Li, Honghui and Wang, Rui and Tian, Xuetao and Liang, Ci and Yan, Fei and Zhang, Junwen and Liu, Zhen},
	year         = 2024,
	journal      = {Software Testing, Verification and Reliability},
	publisher    = {Wiley Online Library},
	pages        = {e1894}
}

@book{schutze2008introduction,
	title        = {Introduction to information retrieval},
	author       = {Sch{\"u}tze, Hinrich and Manning, Christopher D and Raghavan, Prabhakar},
	year         = 2008,
	publisher    = {Cambridge University Press Cambridge},
	volume       = 39
}

@inproceedings{deepatash,
	title        = {DeepAtash: Focused Test Generation for Deep Learning Systems},
	author       = {Tahereh Zohdinasab and Vincenzo Riccio and Paolo Tonella},
	year         = 2023,
	booktitle    = {Proceedings of the 32nd {ACM} {SIGSOFT} International Symposium on Software Testing and Analysis, {ISSTA} 2023, Seattle, WA, USA, July 17-21, 2023},
	publisher    = {{ACM}},
	pages        = {954--966},
	doi          = {10.1145/3597926.3598109},
	url          = {https://doi.org/10.1145/3597926.3598109},
	editor       = {Ren{\'{e}} Just and Gordon Fraser},
	bibsource    = {dblp computer science bibliography, https://dblp.org}
}

@book{wu2012advances,
	title        = {Advances in K-means clustering: a data mining thinking},
	author       = {Wu, Junjie},
	year         = 2012,
	publisher    = {Springer Science \& Business Media}
}

@article{vilone2020explainable,
	title        = {Explainable artificial intelligence: a systematic review},
	author       = {Vilone, Giulia and Longo, Luca},
	year         = 2020,
	journal      = {arXiv preprint arXiv:2006.00093}
}

@inproceedings{dovsilovic2018explainable,
	title        = {Explainable artificial intelligence: A survey},
	author       = {Do{\v{s}}ilovi{\'c}, Filip Karlo and Br{\v{c}}i{\'c}, Mario and Hlupi{\'c}, Nikica},
	year         = 2018,
	booktitle    = {2018 41st International convention on information and communication technology, electronics and microelectronics (MIPRO)},
	pages        = {0210--0215},
	organization = {IEEE}
}

@article{gunning2019xai,
	title        = {XAI—Explainable artificial intelligence},
	author       = {Gunning, David and Stefik, Mark and Choi, Jaesik and Miller, Timothy and Stumpf, Simone and Yang, Guang-Zhong},
	year         = 2019,
	journal      = {Science robotics},
	publisher    = {American Association for the Advancement of Science},
	volume       = 4,
	number       = 37,
	pages        = {eaay7120}
}

@article{simonyan2013deep,
	title        = {Deep inside convolutional networks: Visualising image classification models and saliency maps},
	author       = {Simonyan, Karen and Vedaldi, Andrea and Zisserman, Andrew},
	year         = 2013,
	journal      = {arXiv preprint arXiv:1312.6034}
}

@article{smilkov2017smoothgrad,
	title        = {Smoothgrad: removing noise by adding noise},
	author       = {Smilkov, Daniel and Thorat, Nikhil and Kim, Been and Vi{\'e}gas, Fernanda and Wattenberg, Martin},
	year         = 2017,
	journal      = {arXiv preprint arXiv:1706.03825},
	eprint       = {1706.03825},
	archiveprefix = {arXiv},
	primaryclass = {cs.LG}
}

@inproceedings{selvaraju2017grad,
	title        = {Grad-cam: Visual explanations from deep networks via gradient-based localization},
	author       = {Selvaraju, Ramprasaath R and Cogswell, Michael and Das, Abhishek and Vedantam, Ramakrishna and Parikh, Devi and Batra, Dhruv},
	year         = 2017,
	booktitle    = {Proceedings of the IEEE international conference on computer vision},
	pages        = {618--626}
}

@article{adebayo2018sanity,
	title        = {Sanity checks for saliency maps},
	author       = {Adebayo, Julius and Gilmer, Justin and Muelly, Michael and Goodfellow, Ian and Hardt, Moritz and Kim, Been},
	year         = 2018,
	journal      = {Advances in neural information processing systems},
	volume       = 31
}

@article{Wilcoxon1945,
	title        = {Individual Comparisons by Ranking Methods},
	author       = {Frank Wilcoxon},
	year         = 1945,
	month        = dec,
	journal      = {Biometrics Bulletin},
	publisher    = {{JSTOR}},
	volume       = 1,
	number       = 6,
	pages        = 80,
	doi          = {10.2307/3001968},
	url          = {https://doi.org/10.2307/3001968}
}

@book{cohen1988statistical,
	title        = {Statistical power analysis for the behavioral sciences},
	author       = {Cohen, Jacob},
	year         = 1988,
	publisher    = {L. Erlbaum Associates},
	address      = {Hillsdale, N.J},
	isbn         = {978-1-134-74270-7}
}

@inproceedings{2021-Riccio-ASE,
	title        = {DeepMetis: Augmenting a Deep Learning Test Set to Increase its Mutation Score},
	author       = {Vincenzo Riccio and Nargiz Humbatova and Gunel Jahangirova and Paolo Tonella},
	year         = 2021,
	booktitle    = {Proceedings of the 36th IEEE/ACM International Conference on Automated Software Engineering},
	publisher    = {IEEE/ACM},
	series       = {ASE '21}
}

@inproceedings{2020-Humbatova-ICSE,
	title        = {Taxonomy of Real Faults in Deep Learning Systems},
	author       = {Nargiz Humbatova and Gunel Jahangirova and Gabriele Bavota and Vincenzo Riccio and Andrea Stocco and Paolo Tonella},
	year         = 2020,
	booktitle    = {Proceedings of 42nd International Conference on Software Engineering},
	location     = {Seoul, Republic of Korea},
	publisher    = {ACM},
	address      = {New York, NY, USA},
	series       = {ICSE '20},
	pages        = {12 pages},
	doi          = {10.1145/3377811.3380395},
	isbn         = {978-1-4503-7121-6/20/05},
	numpages     = 11,
	acmid        = 3380395
}

@inproceedings{2025-Baresi-ICSE,
	title        = {Efficient Domain Augmentation for Autonomous Driving Testing Using Diffusion Models},
	author       = {Luciano Baresi and Davide Yi Xian Hu and Andrea Stocco and Paolo Tonella},
	year         = 2025,
	booktitle    = {Proceedings of 47th International Conference on Software Engineering},
	publisher    = {IEEE},
	series       = {ICSE '25},
	abbr         = {ICSE}
}

@article{Zhao-nature,
	title        = {Learning from Longitudinal Data in Electronic Health Record and Genetic Data to Improve Cardiovascular Event Prediction},
	author       = {Zhao, Juan and Feng, QiPing and Wu, Patrick and Lupu, Roxana A. and Wilke, Russell A. and Wells, Quinn S. and Denny, Joshua C. and Wei, Wei-Qi},
	year         = 2019,
	journal      = {Scientific Reports},
	volume       = 9,
	number       = 1,
	pages        = 717,
	doi          = {10.1038/s41598-018-36745-x},
	isbn         = {2045-2322},
	url          = {https://doi.org/10.1038/s41598-018-36745-x},
	id           = {Zhao2019},
	ty           = {JOUR}
}

@article{2020-Riccio-EMSE,
	title        = {{Testing Machine Learning based Systems: A Systematic Mapping}},
	author       = {Vincenzo Riccio and Gunel Jahangirova and Andrea Stocco and Nargiz Humbatova and Michael Weiss and Paolo Tonella},
	year         = 2020,
	journal      = {Empirical Software Engineering},
	publisher    = {Springer}
}

@inproceedings{Xie-ISSTA-2019,
	title        = {DeepHunter: A Coverage-guided Fuzz Testing Framework for Deep Neural Networks},
	author       = {Xie, Xiaofei and Ma, Lei and Juefei-Xu, Felix and Xue, Minhui and Chen, Hongxu and Liu, Yang and Zhao, Jianjun and Li, Bo and Yin, Jianxiong and See, Simon},
	year         = 2019,
	booktitle    = {Proceedings of the 28th ACM SIGSOFT International Symposium on Software Testing and Analysis},
	location     = {Beijing, China},
	publisher    = {ACM},
	address      = {New York, NY, USA},
	series       = {ISSTA '19},
	pages        = {146--157},
	doi          = {10.1145/3293882.3330579},
	isbn         = {978-1-4503-6224-5},
	url          = {http://doi.acm.org/10.1145/3293882.3330579},
	acmid        = 3330579,
	numpages     = 12
}

@inproceedings{2020-Riccio-FSE,
	title        = {{Model-Based Exploration of the Frontier of Behaviours for Deep Learning System Testing}},
	author       = {Vincenzo Riccio and Paolo Tonella},
	year         = 2020,
	booktitle    = {Proceedings of ACM Joint European Software Engineering Conference and Symposium on the Foundations of Software Engineering},
	series       = {ESEC/FSE '20}
}

@article{Julian,
	title        = {Deep Neural Network Compression for Aircraft Collision Avoidance Systems},
	author       = {Kyle D. Julian and Mykel J. Kochenderfer and Michael P. Owen},
	year         = 2018,
	journal      = {CoRR},
	volume       = {abs/1810.04240},
	archiveprefix = {arXiv},
	eprint       = {1810.04240}
}

@inproceedings{2020-Stocco-ICSE,
	title        = {Misbehaviour Prediction for Autonomous Driving Systems},
	author       = {Andrea Stocco and Michael Weiss and Marco Calzana and Paolo Tonella},
	year         = 2020,
	booktitle    = {Proceedings of 42nd International Conference on Software Engineering},
	publisher    = {ACM},
	series       = {ICSE '20},
	pages        = {12 pages}
}

@inproceedings{imbd,
	title        = {Learning Word Vectors for Sentiment Analysis},
	author       = {Maas, Andrew L. and Daly, Raymond E. and Pham, Peter T. and Huang, Dan and Ng, Andrew Y. and Potts, Christopher},
	year         = 2011,
	booktitle    = {Proceedings of the 49th Annual Meeting of the Association for Computational Linguistics: Human Language Technologies - Volume 1},
	location     = {Portland, Oregon},
	publisher    = {Association for Computational Linguistics},
	address      = {Stroudsburg, PA, USA},
	series       = {HLT '11},
	pages        = {142--150},
	isbn         = {978-1-932432-87-9},
	url          = {http://dl.acm.org/citation.cfm?id=2002472.2002491},
	acmid        = 2002491,
	numpages     = 9
}

@inproceedings{lime,
	title        = {"Why Should {I} Trust You?": Explaining the Predictions of Any Classifier},
	author       = {Marco Tulio Ribeiro and Sameer Singh and Carlos Guestrin},
	year         = 2016,
	booktitle    = {Proceedings of the 22nd {ACM} {SIGKDD} International Conference on Knowledge Discovery and Data Mining, San Francisco, CA, USA, August 13-17, 2016},
	pages        = {1135--1144}
}

@article{samek2017explainable,
	title        = {Explainable artificial intelligence: Understanding, visualizing and interpreting deep learning models},
	author       = {Samek, Wojciech and Wiegand, Thomas and M{\"u}ller, Klaus-Robert},
	year         = 2017,
	journal      = {arXiv preprint arXiv:1708.08296}
}

@book{samek2019explainable,
	title        = {Explainable AI: interpreting, explaining and visualizing deep learning},
	author       = {Samek, Wojciech and Montavon, Gr{\'e}goire and Vedaldi, Andrea and Hansen, Lars Kai and M{\"u}ller, Klaus-Robert},
	year         = 2019,
	publisher    = {Springer Nature},
	volume       = 11700
}

@article{jetley2018learn,
	title        = {Learn to pay attention},
	author       = {Jetley, Saumya and Lord, Nicholas A and Lee, Namhoon and Torr, Philip HS},
	year         = 2018,
	journal      = {arXiv preprint arXiv:1804.02391}
}

@article{DBLP:journals/corr/abs-2002-00863,
	title        = {Supporting {DNN} Safety Analysis and Retraining through Heatmap-based Unsupervised Learning},
	author       = {Hazem M. Fahmy and Mojtaba Bagherzadeh and Fabrizio Pastore and Lionel C. Briand},
	year         = 2020,
	journal      = {CoRR},
	volume       = {abs/2002.00863},
	url          = {https://arxiv.org/abs/2002.00863},
	eprinttype   = {arXiv},
	eprint       = {2002.00863}
}

@inproceedings{kim2019guiding,
	title        = {Guiding deep learning system testing using surprise adequacy},
	author       = {Kim, Jinhan and Feldt, Robert and Yoo, Shin},
	year         = 2019,
	booktitle    = {2019 IEEE/ACM 41st International Conference on Software Engineering (ICSE)},
	location     = {Montreal, Quebec, Canada},
	publisher    = {IEEE Press},
	address      = {Piscataway, NJ, USA},
	series       = {ICSE '19},
	pages        = {1039--1049},
	doi          = {10.1109/ICSE.2019.00108},
	url          = {https://doi.org/10.1109/ICSE.2019.00108},
	organization = {IEEE},
	acmid        = 3339634,
	numpages     = 11
}

@article{goodfellow2020generative,
	title        = {Generative adversarial networks},
	author       = {Goodfellow, Ian and Pouget-Abadie, Jean and Mirza, Mehdi and Xu, Bing and Warde-Farley, David and Ozair, Sherjil and Courville, Aaron and Bengio, Yoshua},
	year         = 2020,
	journal      = {Communications of the ACM},
	publisher    = {ACM New York, NY, USA},
	volume       = 63,
	number       = 11,
	pages        = {139--144}
}

@article{doersch2016tutorial,
	title        = {Tutorial on variational autoencoders},
	author       = {Doersch, Carl},
	year         = 2016,
	journal      = {arXiv preprint arXiv:1606.05908}
}

@article{goodfellow2014generative,
	title        = {Generative adversarial nets},
	author       = {Goodfellow, Ian and Pouget-Abadie, Jean and Mirza, Mehdi and Xu, Bing and Warde-Farley, David and Ozair, Sherjil and Courville, Aaron and Bengio, Yoshua},
	year         = 2014,
	journal      = {Advances in neural information processing systems},
	booktitle    = {Advances in Neural Information Processing Systems 27},
	publisher    = {Curran Associates, Inc.},
	volume       = 27,
	pages        = {2672--2680},
	url          = {http://papers.nips.cc/paper/5423-generative-adversarial-nets.pdf},
	editor       = {Z. Ghahramani and M. Welling and C. Cortes and N. D. Lawrence and K. Q. Weinberger}
}

@inproceedings{zohdinasab2021deephyperion,
	title        = {Deephyperion: exploring the feature space of deep learning-based systems through illumination search},
	author       = {Zohdinasab, Tahereh and Riccio, Vincenzo and Gambi, Alessio and Tonella, Paolo},
	year         = 2021,
	booktitle    = {Proceedings of the 30th ACM SIGSOFT International Symposium on Software Testing and Analysis},
	publisher    = {Association for Computing Machinery},
	series       = {ISSTA '21},
	pages        = {79--90}
}

@inproceedings{pei2017deepxplore,
	title        = {Deepxplore: Automated whitebox testing of deep learning systems},
	author       = {Pei, Kexin and Cao, Yinzhi and Yang, Junfeng and Jana, Suman},
	year         = 2017,
	month        = {oct},
	journal      = {Commun. ACM},
	booktitle    = {proceedings of the 26th Symposium on Operating Systems Principles},
	location     = {Shanghai, China},
	publisher    = {ACM},
	address      = {New York, NY, USA},
	series       = {SOSP '17},
	volume       = 62,
	number       = 11,
	pages        = {1--18},
	doi          = {10.1145/3132747.3132785},
	isbn         = {978-1-4503-5085-3},
	issn         = {0001-0782},
	url          = {http://doi.acm.org/10.1145/3132747.3132785},
	acmid        = 3132785,
	numpages     = 18,
	issue_date   = {November 2019}
}

@inproceedings{guo2018dlfuzz,
	title        = {Dlfuzz: Differential fuzzing testing of deep learning systems},
	author       = {Guo, Jianmin and Jiang, Yu and Zhao, Yue and Chen, Quan and Sun, Jiaguang},
	year         = 2018,
	booktitle    = {Proceedings of the 2018 26th ACM Joint Meeting on European Software Engineering Conference and Symposium on the Foundations of Software Engineering},
	pages        = {739--743}
}

@inproceedings{kang2020sinvad,
	title        = {Sinvad: Search-based image space navigation for dnn image classifier test input generation},
	author       = {Kang, Sungmin and Feldt, Robert and Yoo, Shin},
	year         = 2020,
	booktitle    = {Proceedings of the IEEE/ACM 42nd International Conference on Software Engineering Workshops},
	pages        = {521--528}
}

@article{DBLP:journals/corr/abs-2001-11055,
	title        = {Semantic Adversarial Perturbations using Learnt Representations},
	author       = {Isaac Dunn and Tom Melham and Daniel Kroening},
	year         = 2020,
	journal      = {CoRR},
	volume       = {abs/2001.11055},
	url          = {https://arxiv.org/abs/2001.11055},
	eprinttype   = {arXiv},
	eprint       = {2001.11055},
	bibsource    = {dblp computer science bibliography, https://dblp.org}
}

@inproceedings{dunn2021exposing,
	title        = {Exposing previously undetectable faults in deep neural networks},
	author       = {Dunn, Isaac and Pouget, Hadrien and Kroening, Daniel and Melham, Tom},
	year         = 2021,
	booktitle    = {Proceedings of the 30th ACM SIGSOFT International Symposium on Software Testing and Analysis},
	pages        = {56--66}
}

@inproceedings{dola2024cit4dnn,
	title        = {CIT4DNN: Generating Diverse and Rare Inputs for Neural Networks Using Latent Space Combinatorial Testing},
	author       = {Dola, Swaroopa and McDaniel, Rory and Dwyer, Matthew B and Soffa, Mary Lou},
	year         = 2024,
	booktitle    = {Proceedings of the IEEE/ACM 46th International Conference on Software Engineering},
	pages        = {1--13}
}

@book{larman1998applying,
	title        = {Applying UML and patterns},
	author       = {Larman, Craig and others},
	year         = 1998,
	publisher    = {Prentice Hall Upper Saddle River},
	volume       = 2
}

@article{DBLP:journals/corr/MouretC15,
	title        = {Illuminating search spaces by mapping elites},
	author       = {Jean{-}Baptiste Mouret and Jeff Clune},
	year         = 2015,
	journal      = {CoRR},
	volume       = {abs/1504.04909},
	url          = {http://arxiv.org/abs/1504.04909},
	eprinttype   = {arXiv},
	eprint       = {1504.04909},
	bibsource    = {dblp computer science bibliography, https://dblp.org}
}

@article{kingma2013auto,
	title        = {Auto-encoding variational bayes},
	author       = {Kingma, Diederik P and Welling, Max},
	year         = 2013,
	journal      = {arXiv preprint arXiv:1312.6114}
}

@article{cit605761,
	title        = {The AETG system: an approach to testing based on combinatorial design},
	author       = {Cohen, D.M. and Dalal, S.R. and Fredman, M.L. and Patton, G.C.},
	year         = 1997,
	journal      = {IEEE Transactions on Software Engineering},
	volume       = 23,
	number       = 7,
	pages        = {437--444},
	doi          = {10.1109/32.605761}
}

@inproceedings{lambertenghi_ICST25,
	title        = {Benchmarking Image Perturbations for Testing Automated Driving Assistance Systems},
	author       = {Stefano Carlo Lambertenghi and Hannes Leonhard and Andrea Stocco},
	year         = 2025,
	booktitle    = {Proc. of  18th IEEE International Conference on Software Testing, Verification and Validation},
	publisher    = {IEEE},
	series       = {ICST '25}
}

@misc{replication-package,
	title        = {Replication Package.},
	year         = 2025,
	key          = {rep-package},
	howpublished = {\url{https://github.com/ast-fortiss-tum/XMutant}}
}

@article{isa,
	title        = {Identifying and Explaining Safety-critical Scenarios for Autonomous Vehicles via Key Features},
	author       = {Neelofar, Neelofar and Aleti, Aldeida},
	year         = 2024,
	journal      = {ACM Transactions on Software Engineering and Methodology},
	publisher    = {ACM New York, NY},
	volume       = 33,
	number       = 4,
	pages        = {1--32}
}

@article{arxiv.2204.00480,
	title        = {Simulator-based explanation and debugging of hazard-triggering events in DNN-based safety-critical systems},
	author       = {Fahmy, Hazem and Pastore, Fabrizio and Briand, Lionel and Stifter, Thomas},
	year         = 2023,
	journal      = {ACM Transactions on Software Engineering and Methodology},
	publisher    = {ACM New York, NY, USA},
	volume       = 32,
	number       = 4,
	pages        = {1--47},
	doi          = {10.48550/ARXIV.2204.00480},
	url          = {https://arxiv.org/abs/2204.00480},
	copyright    = {arXiv.org perpetual, non-exclusive license}
}

@article{lecun1998gradient,
	title        = {Gradient-based learning applied to document recognition},
	author       = {LeCun, Yann and Bottou, L{\'e}on and Bengio, Yoshua and Haffner, Patrick},
	year         = 1998,
	journal      = {Proceedings of the IEEE},
	publisher    = {Ieee},
	volume       = 86,
	number       = 11,
	pages        = {2278--2324}
}

@misc{unity,
	title        = {Unity3D.},
	year         = 2021,
	key          = {unity},
	howpublished = {\url{https://unity.com}}
}

@article{abs-2201-00009,
	title        = {Improving Deep Neural Network Classification Confidence using Heatmap-based eXplainable {AI}},
	author       = {Erico Tjoa and Hong Jing Khok and Tushar Chouhan and Cuntai Guan},
	year         = 2022,
	journal      = {CoRR},
	volume       = {abs/2201.00009},
	url          = {https://arxiv.org/abs/2201.00009},
	eprinttype   = {arXiv},
	eprint       = {2201.00009}
}

@inproceedings{Abdessalem-ICSE18,
	title        = {Testing Vision-Based Control Systems Using Learnable Evolutionary Algorithms},
	author       = {R. {Ben Abdessalem} and S. {Nejati} and L. {C. Briand} and T. {Stifter}},
	year         = 2018,
	month        = may,
	booktitle    = {2018 IEEE/ACM 40th International Conference on Software Engineering (ICSE)},
	pages        = {1016--1026},
	doi          = {10.1145/3180155.3180160}
}

@inproceedings{Abdessalem-ASE18-1,
	title        = {Testing Autonomous Cars for Feature Interaction Failures Using Many-objective Search},
	author       = {Abdessalem, Raja Ben and Panichella, Annibale and Nejati, Shiva and Briand, Lionel C. and Stifter, Thomas},
	year         = 2018,
	booktitle    = {Proceedings of the 33rd ACM/IEEE International Conference on Automated Software Engineering},
	location     = {Montpellier, France},
	publisher    = {ACM},
	address      = {New York, NY, USA},
	series       = {ASE 2018},
	pages        = {143--154},
	doi          = {10.1145/3238147.3238192},
	isbn         = {978-1-4503-5937-5},
	url          = {http://doi.acm.org/10.1145/3238147.3238192},
	acmid        = 3238192,
	numpages     = 12
}

@article{10.1007/s10515-021-00310-0,
	title        = {DeepGuard: A Framework for Safeguarding Autonomous Driving Systems from Inconsistent Behaviour},
	author       = {Hussain, Manzoor and Ali, Nazakat and Hong, Jang-Eui},
	year         = 2022,
	month        = {may},
	journal      = {Automated Software Engg.},
	publisher    = {Kluwer Academic Publishers},
	address      = {USA},
	volume       = 29,
	number       = 1,
	doi          = {10.1007/s10515-021-00310-0},
	issn         = {0928-8910},
	url          = {https://doi.org/10.1007/s10515-021-00310-0},
	issue_date   = {May 2022},
	numpages     = 32
}

@article{MULLINS2018197,
	title        = {Adaptive generation of challenging scenarios for testing and evaluation of autonomous vehicles},
	author       = {Galen E. Mullins and Paul G. Stankiewicz and R. Chad Hawthorne and Satyandra K. Gupta},
	year         = 2018,
	journal      = {Journal of Systems and Software},
	volume       = 137,
	pages        = {197--215},
	doi          = {https://doi.org/10.1016/j.jss.2017.10.031},
	issn         = {0164-1212},
	url          = {http://www.sciencedirect.com/science/article/pii/S0164121217302546}
}

@inproceedings{Abdessalem-ASE18-2,
	title        = {Testing advanced driver assistance systems using multi-objective search and neural networks},
	author       = {R. {Ben Abdessalem} and S. {Nejati} and L. C. {Briand} and T. {Stifter}},
	year         = 2016,
	month        = sep,
	booktitle    = {2016 31st IEEE/ACM International Conference on Automated Software Engineering (ASE)},
	volume       = {},
	number       = {},
	pages        = {63--74},
	doi          = {},
	issn         = {}
}

@inproceedings{Gambi:2019:ATS:3293882.3330566,
	title        = {Automatically Testing Self-driving Cars with Search-based Procedural Content Generation},
	author       = {Gambi, Alessio and Mueller, Marc and Fraser, Gordon},
	year         = 2019,
	booktitle    = {Proceedings of the 28th ACM SIGSOFT International Symposium on Software Testing and Analysis},
	location     = {Beijing, China},
	publisher    = {ACM},
	address      = {New York, NY, USA},
	series       = {ISSTA 2019},
	pages        = {318--328},
	doi          = {10.1145/3293882.3330566},
	isbn         = {978-1-4503-6224-5},
	url          = {http://doi.acm.org/10.1145/3293882.3330566},
	acmid        = 3330566,
	numpages     = 11
}

@inproceedings{deeproad,
	title        = {DeepRoad: GAN-based Metamorphic Testing and Input Validation Framework for Autonomous Driving Systems},
	author       = {Zhang, Mengshi and Zhang, Yuqun and Zhang, Lingming and Liu, Cong and Khurshid, Sarfraz},
	year         = 2018,
	booktitle    = {Proceedings of the 33rd ACM/IEEE International Conference on Automated Software Engineering},
	location     = {Montpellier, France},
	publisher    = {ACM},
	address      = {New York, NY, USA},
	series       = {ASE 2018},
	pages        = {132--142},
	doi          = {10.1145/3238147.3238187},
	isbn         = {978-1-4503-5937-5},
	url          = {http://doi.acm.org/10.1145/3238147.3238187},
	acmid        = 3238187,
	numpages     = 11
}

@inproceedings{deeptest,
	title        = {DeepTest: Automated Testing of Deep-neural-network-driven Autonomous Cars},
	author       = {Tian, Yuchi and Pei, Kexin and Jana, Suman and Ray, Baishakhi},
	year         = 2018,
	booktitle    = {Proceedings of the 40th International Conference on Software Engineering},
	location     = {Gothenburg, Sweden},
	publisher    = {ACM},
	address      = {New York, NY, USA},
	series       = {ICSE '18},
	pages        = {303--314},
	doi          = {10.1145/3180155.3180220},
	isbn         = {978-1-4503-5638-1},
	url          = {http://doi.acm.org/10.1145/3180155.3180220},
	acmid        = 3180220,
	numpages     = 12
}

@misc{udacity-simulator,
	title        = {{A self-driving car simulator built with Unity}},
	author       = {{Udacity}},
	year         = 2017,
	note         = {Online; accessed 18 August 2019},
	howpublished = {\url{https://github.com/udacity/self-driving-car-sim}}
}

@inproceedings{zhang2018medical,
	title        = {Medical image synthesis with generative adversarial networks for tissue recognition},
	author       = {Zhang, Qianqian and Wang, Haifeng and Lu, Hongya and Won, Daehan and Yoon, Sang Won},
	year         = 2018,
	booktitle    = {2018 IEEE International Conference on Healthcare Informatics (ICHI)},
	organization = {IEEE}
}

@inproceedings{2021-Jahangirova-ICST,
	title        = {Quality Metrics and Oracles for Autonomous Vehicles Testing},
	author       = {Gunel Jahangirova and Andrea Stocco and Paolo Tonella},
	year         = 2021,
	booktitle    = {Proceedings of 14th IEEE International Conference on Software Testing, Verification and Validation},
	publisher    = {IEEE},
	series       = {ICST '21}
}

@misc{05418,
	title        = {VisualBackProp: efficient visualization of CNNs},
	author       = {Bojarski, Mariusz and Choromanska, Anna and Choromanski, Krzysztof and Firner, Bernhard and Jackel, Larry and Muller, Urs and Zieba, Karol},
	year         = 2016,
	publisher    = {arXiv},
	doi          = {10.48550/ARXIV.1611.05418},
	url          = {https://arxiv.org/abs/1611.05418},
	copyright    = {arXiv.org perpetual, non-exclusive license}
}

@article{2023-Stocco-EMSE,
	title        = {Model vs System Level Testing of Autonomous Driving Systems: A Replication and Extension Study},
	author       = {Stocco, Andrea and Pulfer, Brian and Tonella, Paolo},
	year         = 2023,
	month        = {may},
	journal      = {Empirical Softw. Engg.},
	publisher    = {Kluwer Academic Publishers},
	address      = {USA},
	volume       = 28,
	number       = 3,
	pages        = 73,
	doi          = {10.1007/s10664-023-10306-x},
	issn         = {1382-3256},
	url          = {https://doi.org/10.1007/s10664-023-10306-x},
	issue_date   = {May 2023},
	numpages     = 26
}

@misc{10631,
	title        = {Interpretable Learning for Self-Driving Cars by Visualizing Causal Attention},
	author       = {Kim, Jinkyu and Canny, John},
	year         = 2017,
	publisher    = {arXiv},
	doi          = {10.48550/ARXIV.1703.10631},
	url          = {https://arxiv.org/abs/1703.10631},
	copyright    = {arXiv.org perpetual, non-exclusive license}
}

@misc{09405,
	title        = {Explainable Object-induced Action Decision for Autonomous Vehicles},
	author       = {Xu, Yiran and Yang, Xiaoyin and Gong, Lihang and Lin, Hsuan-Chu and Wu, Tz-Ying and Li, Yunsheng and Vasconcelos, Nuno},
	year         = 2020,
	publisher    = {arXiv},
	doi          = {10.48550/ARXIV.2003.09405},
	url          = {https://arxiv.org/abs/2003.09405},
	copyright    = {arXiv.org perpetual, non-exclusive license}
}

@misc{iso26262,
	title        = {{Road vehicles -- Functional safety}},
	author       = {ISO},
	year         = {{2011}},
	publisher    = {{ISO, Geneva, Switzerland}},
	number       = {{ISO 26262}},
	key          = {{ISO 26262}},
	type         = {{Norm}}
}

@inproceedings{2022-Stocco-ASE,
	title        = {{ThirdEye: Attention Maps for Safe Autonomous Driving Systems}},
	author       = {Andrea Stocco and Paulo J. Nunes and Marcelo d'Amorim and Paolo Tonella},
	year         = 2022,
	booktitle    = {Proceedings of 37th IEEE/ACM International Conference on Automated Software Engineering},
	publisher    = {IEEE/ACM},
	series       = {ASE '22},
	doi          = {10.1145/3551349.3556968},
	note         = {Accepted},
	abbr         = {ASE}
}

@article{survey-lei-ma,
	title        = {A Survey on Automated Driving System Testing: Landscapes and Trends},
	author       = {Shuncheng Tang and Zhenya Zhang and Yi Zhang and Jixiang Zhou and Yan Guo and Shuang Liu and Shengjian Guo and Yan{-}Fu Li and Lei Ma and Yinxing Xue and Yang Liu},
	year         = 2022,
	journal      = {CoRR},
	publisher    = {arXiv},
	volume       = {abs/2206.05961},
	doi          = {10.48550/arXiv.2206.05961},
	url          = {https://arxiv.org/abs/2206.05961},
	copyright    = {arXiv.org perpetual, non-exclusive license},
	eprinttype   = {arXiv},
	eprint       = {2206.05961},
	bibsource    = {dblp computer science bibliography, https://dblp.org}
}

@article{Fleiss:1971,
	title        = {Measuring nominal scale agreement among many raters.},
	author       = {Fleiss, Joseph L},
	year         = 1971,
	journal      = {Psychological bulletin},
	publisher    = {American Psychological Association},
	volume       = 76,
	number       = 5,
	pages        = 378
}

@article{Landis:1977,
	title        = {The measurement of observer agreement for categorical data},
	author       = {Landis, J Richard and Koch, Gary G},
	year         = 1977,
	journal      = {Biometrics},
	publisher    = {JSTOR},
	pages        = {159--174}
}

@inproceedings{2023-Riccio-ICSE,
	title        = {When and Why Test Generators for Deep Learning Produce Invalid Inputs: an Empirical Study},
	author       = {Vincenzo Riccio and Paolo Tonella},
	year         = 2023,
	booktitle    = {Proceedings of 45th International Conference on Software Engineering},
	publisher    = {ACM},
	series       = {ICSE '23},
	pages        = {12 pages},
	abbr         = {ICSE}
}

@article{mnist,
	title        = {The mnist database of handwritten digit images for machine learning research},
	author       = {Deng, Li},
	year         = 2012,
	journal      = {IEEE Signal Processing Magazine},
	publisher    = {IEEE},
	volume       = 29,
	number       = 6,
	pages        = {141--142}
}

@article{neural_coverage_new,
	title        = {You Can't See the Forest for Its Trees: Assessing Deep Neural Network Testing via NeuraL Coverage},
	author       = {Yuanyuan Yuan and Qi Pang and Shuai Wang},
	year         = 2021,
	journal      = {CoRR},
	volume       = {abs/2112.01955},
	url          = {https://arxiv.org/abs/2112.01955},
	eprinttype   = {arXiv},
	eprint       = {2112.01955},
	bibsource    = {dblp computer science bibliography, https://dblp.org}
}

@article{2024-Biagiola-EMSE,
	title        = {Two is Better Than One: Digital Siblings to Improve Autonomous Driving Testing},
	author       = {Matteo Biagiola and Andrea Stocco and Vincenzo Riccio and Paolo Tonella},
	year         = 2024,
	journal      = {Empirical Software Engineering},
	publisher    = {Springer}
}

@article{Jha2019MLBasedFI,
	title        = {ML-Based Fault Injection for Autonomous Vehicles: A Case for Bayesian Fault Injection},
	author       = {Saurabh Jha and Subho Sankar Banerjee and Timothy Tsai and Siva Kumar Sastry Hari and Michael B. Sullivan and Zbigniew T. Kalbarczyk and Stephen W. Keckler and Ravishankar Krishnan Iyer},
	year         = 2019,
	journal      = {2019 49th Annual IEEE/IFIP International Conference on Dependable Systems and Networks (DSN)},
	pages        = {112--124},
	url          = {https://api.semanticscholar.org/CorpusID:195776612}
}

@article{9712397,
	title        = {Learning Configurations of Operating Environment of Autonomous Vehicles to Maximize their Collisions},
	author       = {Lu, Chengjie and Shi, Yize and Zhang, Huihui and Zhang, Man and Wang, Tiexin and Yue, Tao and Ali, Shaukat},
	year         = 2023,
	journal      = {IEEE Transactions on Software Engineering},
	volume       = 49,
	number       = 1,
	pages        = {384--402},
	doi          = {10.1109/TSE.2022.3150788}
}

@misc{zhongETAL2021,
	title        = {Neural Network Guided Evolutionary Fuzzing for Finding Traffic Violations of Autonomous Vehicles},
	author       = {Zhong, Ziyuan and Kaiser, Gail and Ray, Baishakhi},
	year         = 2021,
	publisher    = {arXiv},
	copyright    = {arXiv.org perpetual, non-exclusive license}
}

@inproceedings{liETAL2020,
	title        = {AV-FUZZER: Finding Safety Violations in Autonomous Driving Systems},
	author       = {Li, Guanpeng and Li, Yiran and Jha, Saurabh and Tsai, Timothy and Sullivan, Michael and Hari, Siva Kumar Sastry and Kalbarczyk, Zbigniew and Iyer, Ravishankar},
	year         = 2020,
	booktitle    = {2020 IEEE 31st International Symposium on Software Reliability Engineering (ISSRE)},
	volume       = {},
	number       = {},
	pages        = {25--36},
	doi          = {10.1109/ISSRE5003.2020.00012}
}

@inproceedings{10.1145/3597926.3598072,
	title        = {BehAVExplor: Behavior Diversity Guided Testing for Autonomous Driving Systems},
	author       = {Cheng, Mingfei and Zhou, Yuan and Xie, Xiaofei},
	year         = 2023,
	booktitle    = {Proceedings of the 32nd ACM SIGSOFT International Symposium on Software Testing and Analysis},
	location     = {Seattle, WA, USA},
	publisher    = {Association for Computing Machinery},
	address      = {New York, NY, USA},
	series       = {ISSTA 2023},
	pages        = {488–500},
	doi          = {10.1145/3597926.3598072},
	isbn         = 9798400702211,
	url          = {https://doi.org/10.1145/3597926.3598072},
	numpages     = 13
}

@inproceedings{drivefuzz,
	title        = {{DriveFuzz}},
	author       = {Seulbae Kim and Major Liu and Junghwan "John" Rhee and Yuseok Jeon and Yonghwi Kwon and Chung Hwan Kim},
	year         = 2022,
	month        = {nov},
	booktitle    = {Proceedings of the 2022 {ACM} {SIGSAC} Conference on Computer and Communications Security},
	publisher    = {{ACM}},
	doi          = {10.1145/3548606.3560558},
	url          = {https://doi.org/10.1145%2F3548606.3560558}
}

@inproceedings{10304866,
	title        = {An Empirical Study on Low- and High-Level Explanations of Deep Learning Misbehaviours},
	author       = {Zohdinasab, Tahereh and Riccio, Vincenzo and Tonella, Paolo},
	year         = 2023,
	booktitle    = {2023 ACM/IEEE International Symposium on Empirical Software Engineering and Measurement (ESEM)},
	volume       = {},
	number       = {},
	pages        = {1--11},
	doi          = {10.1109/ESEM56168.2023.10304866}
}

@article{farin1983algorithms,
	title        = {Algorithms for rational B{\'e}zier curves},
	author       = {Farin, Gerald},
	year         = 1983,
	journal      = {Computer-aided design},
	publisher    = {Elsevier},
	volume       = 15,
	number       = 2,
	pages        = {73--77}
}

@incollection{catmull1974class,
	title        = {A class of local interpolating splines},
	author       = {Catmull, Edwin and Rom, Raphael},
	year         = 1974,
	booktitle    = {Computer aided geometric design},
	publisher    = {Elsevier},
	pages        = {317--326}
}

@inproceedings{shrikumar2017learning,
	title        = {Learning important features through propagating activation differences},
	author       = {Shrikumar, Avanti and Greenside, Peyton and Kundaje, Anshul},
	year         = 2017,
	booktitle    = {International conference on machine learning},
	pages        = {3145--3153},
	organization = {PMLR}
}

@inproceedings{li2021experimental,
	title        = {An experimental study of quantitative evaluations on saliency methods},
	author       = {Li, Xiao-Hui and Shi, Yuhan and Li, Haoyang and Bai, Wei and Cao, Caleb Chen and Chen, Lei},
	year         = 2021,
	booktitle    = {Proceedings of the 27th ACM sigkdd conference on knowledge discovery \& data mining},
	pages        = {3200--3208}
}

@misc{zheng2023judgingllmasajudgemtbenchchatbot,
	title        = {Judging LLM-as-a-Judge with MT-Bench and Chatbot Arena},
	author       = {Lianmin Zheng and Wei-Lin Chiang and Ying Sheng and Siyuan Zhuang and Zhanghao Wu and Yonghao Zhuang and Zi Lin and Zhuohan Li and Dacheng Li and Eric P. Xing and Hao Zhang and Joseph E. Gonzalez and Ion Stoica},
	year         = 2023,
	url          = {https://arxiv.org/abs/2306.05685},
	eprint       = {2306.05685},
	archiveprefix = {arXiv},
	primaryclass = {cs.CL}
}

@misc{wang2025llmsreplacehumanevaluators,
	title        = {Can LLMs Replace Human Evaluators? An Empirical Study of LLM-as-a-Judge in Software Engineering},
	author       = {Ruiqi Wang and Jiyu Guo and Cuiyun Gao and Guodong Fan and Chun Yong Chong and Xin Xia},
	year         = 2025,
	url          = {https://arxiv.org/abs/2502.06193},
	eprint       = {2502.06193},
	archiveprefix = {arXiv},
	primaryclass = {cs.SE}
}

@article{sinvad-tosem,
	title        = {Deceiving Humans and Machines Alike: Search-Based Test Input Generation for DNNs Using Variational Autoencoders},
	author       = {Kang, Sungmin and Feldt, Robert and Yoo, Shin},
	year         = 2024,
	month        = dec,
	journal      = {ACM Transactions on Software Engineering Methodologies},
	volume       = 33,
	pages        = {103:1--24},
	doi          = {10.1145/3635706},
	issue        = 4
}

@inproceedings{Ma:2018:DMT:3238147.3238202,
	title        = {DeepGauge: Multi-granularity Testing Criteria for Deep Learning Systems},
	author       = {Ma, Lei and Juefei-Xu, Felix and Zhang, Fuyuan and Sun, Jiyuan and Xue, Minhui and Li, Bo and Chen, Chunyang and Su, Ting and Li, Li and Liu, Yang and Zhao, Jianjun and Wang, Yadong},
	year         = 2018,
	booktitle    = {Proceedings of the 33rd ACM/IEEE International Conference on Automated Software Engineering},
	location     = {Montpellier, France},
	publisher    = {ACM},
	address      = {New York, NY, USA},
	series       = {ASE 2018},
	pages        = {120--131},
	doi          = {10.1145/3238147.3238202},
	isbn         = {978-1-4503-5937-5},
	url          = {http://doi.acm.org/10.1145/3238147.3238202},
	acmid        = 3238202,
	numpages     = 12
}

@inproceedings{2025-Maryam-ICST,
	title        = {{Benchmarking Generative AI Models for Deep Learning Test Input Generation}},
	author       = {Maryam and Matteo Biagiola and Andrea Stocco and Vincenzo Riccio},
	year         = 2025,
	booktitle    = {Proceedings of 18th IEEE International Conference on Software Testing, Verification and Validation},
	publisher    = {IEEE},
	series       = {ICST '25},
	pages        = {12 pages}
}

@misc{zhang2024systematicreviewlongtailedlearning,
	title        = {A Systematic Review on Long-Tailed Learning},
	author       = {Chongsheng Zhang and George Almpanidis and Gaojuan Fan and Binquan Deng and Yanbo Zhang and Ji Liu and Aouaidjia Kamel and Paolo Soda and João Gama},
	year         = 2024,
	url          = {https://arxiv.org/abs/2408.00483},
	eprint       = {2408.00483},
	archiveprefix = {arXiv},
	primaryclass = {cs.LG}
}

@article{Goodfellow2014ExplainingAH,
  title={Explaining and Harnessing Adversarial Examples},
  author={Ian J. Goodfellow and Jonathon Shlens and Christian Szegedy},
  journal={CoRR},
  year={2014},
  volume={abs/1412.6572},
  url={https://api.semanticscholar.org/CorpusID:6706414}
}

@article{nvidia-dave2,
	title        = {End to End Learning for Self-Driving Cars.},
	author       = {Bojarski, Mariusz and Testa, Davide Del and Dworakowski, Daniel and Firner, Bernhard and Flepp, Beat and Goyal, Prasoon and Jackel, Lawrence D. and Monfort, Mathew and Muller, Urs and Zhang, Jiakai and Zhang, Xin and Zhao, Jake and Zieba, Karol},
	year         = 2016,
	journal      = {CoRR},
	volume       = {abs/1604.07316},
	url          = {http://arxiv.org/abs/1604.07316}
}

@article{2023-Stocco-TSE,
	title        = {{Mind the Gap! A Study on the Transferability of Virtual Versus Physical-World Testing of Autonomous Driving Systems}},
	author       = {A. Stocco and B. Pulfer and P. Tonella},
	year         = 2023,
	month        = {apr},
	journal      = {IEEE Transactions on Software Engineering},
	publisher    = {IEEE Computer Society},
	address      = {Los Alamitos, CA, USA},
	volume       = 49,
	number       = {04},
	pages        = {1928--1940},
	doi          = {10.1109/TSE.2022.3202311},
	issn         = {1939-3520}
}

@article{10.1007/s10664-023-10433-5,
	title        = {Evaluating the impact of flaky simulators on testing autonomous driving systems},
	author       = {Amini, Mohammad Hossein and Naseri, Shervin and Nejati, Shiva},
	year         = 2024,
	month        = {feb},
	journal      = {Empirical Softw. Engg.},
	publisher    = {Kluwer Academic Publishers},
	address      = {USA},
	volume       = 29,
	number       = 2,
	doi          = {10.1007/s10664-023-10433-5},
	issn         = {1382-3256},
	url          = {https://doi.org/10.1007/s10664-023-10433-5},
	issue_date   = {Mar 2024},
	numpages     = 30
}

@misc{Guannam_ETAL_FSE22,
	title        = {Testing of Autonomous Driving Systems: Where Are We and Where Should We Go?},
	author       = {Lou, Guannan and Deng, Yao and Zheng, Xi and Zhang, Mengshi and Zhang, Tianyi},
	year         = 2021,
	publisher    = {arXiv},
	doi          = {10.48550/ARXIV.2106.12233},
	url          = {https://arxiv.org/abs/2106.12233},
	copyright    = {arXiv.org perpetual, non-exclusive license}
}

@inproceedings{Grad-CAM++_2018,
	title        = {Grad-CAM++: Generalized Gradient-Based Visual Explanations for Deep Convolutional Networks},
	author       = {Chattopadhay, Aditya and Sarkar, Anirban and Howlader, Prantik and Balasubramanian, Vineeth N},
	year         = 2018,
	month        = mar,
	booktitle    = {2018 IEEE Winter Conference on Applications of Computer Vision (WACV)},
	publisher    = {IEEE},
	doi          = {10.1109/wacv.2018.00097},
	url          = {http://dx.doi.org/10.1109/WACV.2018.00097}
}

@misc{integratedgradients,
	title        = {Axiomatic Attribution for Deep Networks},
	author       = {Mukund Sundararajan and Ankur Taly and Qiqi Yan},
	year         = 2017,
	eprint       = {1703.01365},
	archiveprefix = {arXiv},
	primaryclass = {cs.LG}
}

@article{du2019techniques,
	title        = {Techniques for interpretable machine learning},
	author       = {Du, Mengnan and Liu, Ninghao and Hu, Xia},
	year         = 2019,
	journal      = {Communications of the ACM},
	publisher    = {ACM New York, NY, USA},
	volume       = 63,
	number       = 1,
	pages        = {68--77}
}

@misc{man_sentimental_imdb_lstm,
	title        = {Sentimental Analysis on IMDB by LSTM},
	author       = {Vincent Kao},
	year         = 2018,
	url          = {https://www.kaggle.com/code/vincentman0403/sentimental-analysis-on-imdb-by-lstm/notebook},
	note         = {Accessed: 2024-08-18}
}

@article{hochreiter1997long,
	title        = {Long Short-term Memory},
	author       = {Hochreiter, S},
	year         = 1997,
	journal      = {Neural Computation MIT-Press}
}

@article{miller1995wordnet,
	title        = {WordNet: a lexical database for English},
	author       = {Miller, George A},
	year         = 1995,
	journal      = {Communications of the ACM},
	publisher    = {ACM New York, NY, USA},
	volume       = 38,
	number       = 11,
	pages        = {39--41}
}

@misc{openai,
	title        = {The most powerful platform for building AI products},
	author       = {OpenAI},
	year         = 2024,
	note         = {Accessed: 2024-09-08},
	howpublished = {\url{https://openai.com/api/}}
}

@misc{GPT,
	title        = {GPT-4o mini: advancing cost-efficient intelligence},
	author       = {OpenAI},
	year         = 2024,
	note         = {Accessed: 2024-09-08},
	howpublished = {\url{https://openai.com/index/gpt-4o-mini-advancing-cost-efficient-intelligence/}}
}

@article{shannon1948mathematical,
  title={A mathematical theory of communication},
  author={Shannon, Claude E},
  journal={The Bell system technical journal},
  volume={27},
  number={3},
  pages={379--423},
  year={1948},
  publisher={Nokia Bell Labs}
}

@online{kaggle_vgg16_mnist,
  author       = {Virat Kothari},
  title        = {Image Classification of MNIST using {VGG16}},
  year         = {2020},  
  howpublished = {Kaggle Notebook},
  url          = {https://www.kaggle.com/code/viratkothari/image-classification-of-mnist-using-vgg16}
}

@article{ferjad2020icml,
  title = {Reliable Fidelity and Diversity Metrics for Generative Models},
  author = {Naeem, Muhammad Ferjad and Oh, Seong Joon and Uh, Youngjung and Choi, Yunjey and Yoo, Jaejun},
  year = {2020},
  booktitle = {International Conference on Machine Learning},
}

@INPROCEEDINGS{fongInterpretableExplanations,
  author={Fong, Ruth C. and Vedaldi, Andrea},
  booktitle={2017 IEEE International Conference on Computer Vision (ICCV)}, 
  title={Interpretable Explanations of Black Boxes by Meaningful Perturbation}, 
  year={2017},
  volume={},
  number={},
  pages={3449-3457},
  doi={10.1109/ICCV.2017.371}}

\end{document}